\documentclass[12pt,a4paper]{article}
\usepackage{amsmath,amssymb,amsthm}
\numberwithin{equation}{section}

\usepackage[margin=1.0in]{geometry}
\def\a{\alpha}
\def\b{\beta}

\def\g{\gamma}
\def\d{\delta}
\def\e{\epsilon}
\def\ve{\varepsilon}

\def\br{{\bf r}}
\def\et{\eta}

\def\l{\lambda}
\def\m{\mu}
\def\n{\nu}

\def\r{\rho}

\def\s{\sigma}
\def\t{\tau}

\def\vp{\phi}

\def\ps{\psi}
\def\G{{\mathcal G}}
\def\D{\Delta}

\def\X{{\mathcal X}}
\def\P{\Pi}
\def\S{\Sigma}

\def\P{\Phi}

\def\O{{\mathcal O}}
\def\om{\omega}

\def\RR{\mathbb R}

\def\hp{\h\phi}
\def\Cc{{\mathcal C}}
\def\h{\widehat}

\newcommand{\C}[1]{$(\ref{#1})$}
\newcommand{\hph}[1]{{\hphantom{#1}}}
\def\o{\over}

\def\tria{\triangleright}

\def\hlf{\frac{1}{2}}
\def\lp{\left(}
\def\rp{\right)}

\def\ie{{\it i.e. }}

\makeatletter
\renewcommand\section{\@startsection {section}{1}{\z@}%
                                   {-3.5ex \@plus -1ex \@minus -.2ex}
                                   {2.3ex \@plus.2ex}%
                                   {\normalfont\large\bfseries}}
\renewcommand\subsection{\@startsection{subsection}{2}{\z@}%
                                     {-3.25ex\@plus -1ex \@minus -.2ex}%
                                     {1.5ex \@plus .2ex}%
                                     {\normalfont\bfseries}}
\makeatother

\begin{document}

\begin{center}
\addtolength{\baselineskip}{.5mm}
\thispagestyle{empty}
\begin{flushright}
{\sc MIFPA-14-12}\\
\end{flushright}

\vspace{20mm}

{\Large  \bf The $\a'$ Expansion On A Compact Manifold \vskip.3cm Of Exceptional Holonomy}
\\[15mm]
{Katrin Becker$^1$, Daniel Robbins$^2$, and Edward Witten$^3$}
\\[5mm]
{\it $^1$George P. and Cynthia W. Mitchell Institute for }\\
{\it Fundamental Physics and Astronomy, Texas A\& M University,}\\
{\it College Station, TX 77843-4242, USA}\\[5mm]
{\it $^2$Institute for Theoretical Physics, University of Amsterdam,} \\
{\it Postbus 94485, 1090 GL, Amsterdam, The Netherlands}\\ [5mm]
{\it $^3$School of Natural Sciences,  Institute for Advanced Study,} \\
{\it Einstein Drive, Princeton, NJ 08540 USA}\\
{\rm and} \\
{\it Department of Physics, University of Washington, }
\\ {\it Seattle, Washington  98195 USA }

\vspace{20mm}

{\bf  Abstract}
\end{center}

\def\R{{\Bbb R}}
\def\N{{\mathcal N}}
\def\Spin{\rm{Spin}}
In the  approximation corresponding to the classical Einstein equations, which is valid at large radius, string theory compactification
on a compact manifold $M$ of $G_2$ or $\mathrm{Spin}(7)$ holonomy gives a supersymmetric vacuum in three or two dimensions.
Do $\alpha'$ corrections to the Einstein equations disturb this statement?  Explicitly analyzing the leading correction, we show
that the metric of $M$ can be adjusted to maintain supersymmetry.   Beyond leading order, a general argument based on low energy
effective field theory in spacetime implies that this is true exactly  (not just to all finite orders in $\alpha'$).
A more elaborate field theory argument that includes the massive Kaluza-Klein modes matches the structure found in explicit
calculations.  In M-theory compactification on  a manifold $M$ of $G_2$ or $\Spin(7)$ holonomy, similar results hold to all
orders in the inverse radius of $M$ -- but not exactly.   The classical moduli space of $G_2$ metrics on a manifold $M$ is known
to be locally a Lagrangian submanifold of $H^3(M,\R)\oplus H^4(M,\R)$.  We show that this remains valid to all orders in the
$\alpha'$ or inverse radius expansion.

\vfill
\newpage

\tableofcontents

\section{Introduction}

Let $M$ be a compact seven-manifold of $G_2$ holonomy.  Compactification on $M$ gives a classical solution of ten-dimensional supergravity, with unbroken
supersymmetry in three dimensions.  Is the analogous statement true
in string theory, allowing for $\alpha'$ corrections? Differently put, corresponding to a classical solution of Einstein's equations with $G_2$ holonomy,
is there a superconformally invariant two-dimensional $\sigma$-model with target $M$?

To be more precise, we consider the question in the context of Type II
superstring theory (or somewhat similarly, the heterotic string with the spin connection embedded in the gauge group).  Then the question is whether there
is a  family of $\sigma$-models (without torsion)
with target $M$ and  $(1,1)$ superconformal symmetry, depending on the moduli of the classical $G_2$ metric.
The question arises because \cite{gvz} the $\alpha'$ expansion in $\sigma$-model
perturbation theory generates corrections to the Einstein equations and to the conditions for supersymmetry.  A ``yes'' answer means that the classical
metric of $G_2$ holonomy can be modified to compensate for these corrections.

  This natural question appears not to have been addressed in the literature.  A somewhat different question has been answered \cite{sv}:
  {\it if} such a $\sigma$-model
  exists, what is its chiral algebra?  The answer involves an interesting extension of the $\N=1$ superconformal algebra in two dimensions.   Also,
  the leading $\alpha'$ correction has been analyzed in some examples \cite{Lu:2003ze}.  Explicit formulas were found showing that the
  leading correction does not destroy spacetime supersymmetry in these examples.\footnote{The examples
in question are explicit, complete but non-compact manifolds of $G_2$ holonomy.  In the present paper, we phrase our statements in terms of compact manifolds
  of  exceptional holonomy to avoid some analytical details, but one expects our main results to carry over to a large class of complete but
  not compact examples.}

  In the context of compactification on a six-dimensional Calabi-Yau manifold $X$, there is a superficially similar question:  can a classical metric of $SU(3)$ holonomy
  be corrected to compensate for the modifications of Einstein's equations that arise in $\sigma$-model perturbation theory and so to
  maintain spacetime supersymmetry?
  This question has a nice answer from the point of view of the two-dimensional $\sigma$-model.  One uses the fact that (in a $\sigma$-model
  without torsion\footnote{There are more general $(2,2)$ models with torsion \cite{GHR}, but we need not consider them here.}) the condition for $(2,2)$ worldsheet supersymmetry,
  without assuming conformal invariance, is that the target space $X$ should be Kahler. Moreover, it is possible to regularize the $\sigma$-model preserving
  $(2,2)$ supersymmetry.  Hence in analyzing the renormalization group flow on the metric of $X$
  that is induced by $\sigma$-model corrections, one can consider only flows in the space of Kahler metrics.  At one-loop order, one meets the classical
  Einstein equations, and this is where the Calabi-Yau condition comes in.  What about the higher order corrections? They actually give a flow in the space of Kahler potentials (that is, a flow that keeps fixed the Kahler class of the target space
  metric).  Once one knows this, one can easily argue \cite{NS} that the metric of $X$ can be corrected order by order in $\sigma$-model perturbation theory (and even
  exactly) to maintain superconformal
  invariance and therefore spacetime supersymmetry.\footnote{As a statement about worldsheet superconformal invariance, this argument actually
  works for Calabi-Yau $n$-folds for any $n$.}

  Unfortunately, we have not been able to generalize this argument for a manifold $M$  of $G_2$ holonomy.  Roughly, to do this one would want a global or ``off-shell''
  version of the chiral algebra that was described in \cite{sv}.  In other words, one would want to identify part of this structure (analogous to global $(2,2)$ supersymmetry in the Calabi-Yau case) that can be preserved in the presence of a suitable regulator, so that it is valid in $\sigma$-model perturbation theory.
  For example, it might be that with some regularization, the renormalization group flow of the $\sigma$-model
  with target $M$ takes place only in the space of metrics that can be derived from a  (not necessarily torsion-free\footnote{A $G_2$-structure
  on a seven-manifold $M$  is a three-form $\vp$ that obeys mild inequalities which ensure that the formula (\ref{metric1}) does define
  a Riemannian metric.  The $G_2$-structure is said  to be torsion-free (and in this case $M$ is called a $G_2$ manifold) if the metric
  has $G_2$ holonomy. }) $G_2$-structure derived from a closed three-form $\phi$ with a fixed cohomology class.
  This would be analogous to the fact that
  in the Calabi-Yau case, the flow (with a regularization that preserves global $(2,2)$ supersymmetry) takes place only
  in the space of Kahler forms  with a fixed cohomology class.  Given
  such a statement, perhaps one could imitate the argument in \cite{NS}.  But we have been unable to find such a statement.

  In the Calabi-Yau case,  an alternative argument  uses elementary properties of the spacetime effective action
   to show that $\alpha'$ corrections do not destroy spacetime
  supersymmetry.  A version of the argument\footnote{\label{mention}This argument is not restricted to the case of embedding the spin connection in the gauge group; it
  applies to the larger class of supersymmetric heterotic string compactifications described in \cite{EW}.}
  appropriate to the heterotic string, in which Calabi-Yau compactification preserves $\N=1$ supersymmetry
  in spacetime, proceeds as follows \cite{EW}.
  To disturb $\N=1$ supersymmetry, one must generate a correction either to the spacetime superpotential
 or to the Fayet-Iliopoulos (FI) $D$-terms.  Simple arguments based on scaling and holomorphy show that at string tree level, $\alpha'$ corrections cannot
 generate either of these effects.\footnote{For similar reasons, string loop corrections cannot generate a superpotential \cite{DS,DStwo}, although they can generate
 the FI terms at one-loop order \cite{DSW}.   Nonperturbatively in the string coupling,
 spacetime instantons can and typically do generate a nonperturbative superpotential in these models.
Nonperturbatively in $\alpha'$, the same is true of worldsheet instantons in the general class of models mentioned in footnote \ref{mention}, though not in models constructed
 via the standard embedding of the spin connection in the gauge group.}
 Similar reasoning applies
 to Type II superstring theory on a Calabi-Yau manifold. In this case, one has $\N=2$ supersymmetry in spacetime, which can only be disturbed by FI terms (there is
 no analog of the superpotential), and after disposing of these terms, one learns that spacetime supersymmetry is unbroken in the full quantum string theory.

 In Type II superstring theory on a manifold of $G_2$ holonomy, a similar argument based on holomorphy of the spacetime effective action shows that
 $\sigma$-model corrections cannot disturb spacetime supersymmetry.  Such a model has $\N=2$ supersymmetry in three-dimensional
 spacetime, which is similar to $\N=1$ in four dimensions, and {\it a priori}
 spacetime supersymmetry might be disturbed by a correction to the superpotential or by FI $D$-terms.  Holomorphy ensures that $\alpha'$
 corrections cannot correct the spacetime superpotential:  the superpotential is a function of chiral superfields whose imaginary parts are RR fields or axion-like
 modes from the NS-NS sector, all of which decouple at zero momentum in $\sigma$-model perturbation theory.  And since the gauge fields of Type II
 superstring theory on a $G_2$ manifold arise from the RR sector, their FI terms would actually violate a symmetry of $\sigma$-model perturbation theory
 (the symmetry $(-1)^{F_L}$ that counts left-moving fermions mod 2).

 The reason for the present paper is that although we consider the argument in the last paragraph to be satisfactory, we wanted to understand
 what happens more explicitly.  We begin in section \ref{gtwom} by considering the first non-trivial $\alpha'$ correction to the Einstein equations and the
 conditions for unbroken supersymmetry in Type II compactification on a $G_2$ manifold.  We show explicitly that the metric can always be corrected to maintain
 spacetime supersymmetry in this order.    We find that the same will be true in higher orders if a certain four-form $\alpha$ and five-form $\beta$ (which can
 be computed order by order in $\alpha'$) are always
 exact.  We do not know how to show this directly from $\sigma$-model considerations.  However, we show that the spacetime
 arguments mentioned in the last paragraph amount to predicting the exactness of a certain four-form and five-form.  We expect these to coincide with the $\alpha$
 and $\beta$ that come from $\sigma$-model perturbation theory.

An obvious question is to consider instead compactification on an eight-manifold of $\Spin(7)$ holonomy.
 In section \ref{spinseven}, we show explicitly that once again in this case, the leading $\alpha'$ correction to the supersymmetry transformations can be compensated by adjusting
 the metric to maintain spacetime supersymmetry.  Simple arguments based on the spacetime effective action predict that this result
 must persist in higher orders.  For example, Type IIA compactification on a $\Spin(7)$ manifold gives a model with $(1,1)$ supersymmetry in two dimensions.  In this model,
 the symmetry $(-1)^{F_L}$ acts as a discrete $R$-symmetry that -- when combined with the decoupling of RR fields of zero momentum in $\sigma$-model perturbation
 theory -- ensures that $\alpha'$ corrections cannot break spacetime supersymmetry.  We elaborate on such arguments in section \ref{intlow}.

These questions have some obvious further generalizations.
One can consider M-theory compactifications to four or three dimensions on a  $G_2$
 or $\Spin(7)$ manifold $M$.  The analog of the $\alpha'$ expansion is an expansion in powers of $1/\br$, with $\br$ the radius of $M$.  In each of these
 cases, supersymmetry is maintained to all finite orders\footnote{In the $G_2$ case, nonperturbatively in $1/\br$,  instantons
 derived from wrapped M2-branes generate a spacetime superpotential
 \cite{HM} that triggers supersymmetry breaking.} in $1/\br$.  On a  $G_2$ manifold,
 holomorphy of the superpotential together with decoupling of the $C$-field at zero momentum leads to essentially the same argument as in the case of Type
 II superstring theory on a $G_2$ manifold.  On a $\Spin(7)$ manifold, as explained in section \ref{spinseven}, one makes much the same argument as in the
 Type IIA case, using a reflection
 symmetry in the non-compact directions instead of its string theory reduction, which is $(-1)^{F_L}$.

 Finally, one can ask about the heterotic string on a manifold of $G_2$ or $\Spin(7)$ holonomy.  There are in the supergravity limit many supersymmetric
 compactifications that are not obtained by simply embedding the spin connection in the gauge group in the usual fashion.\footnote{Instead, the gauge field
 obeys an appropriate equation associated to spacetime supersymmetry.  On a $G_2$ manifold, this equation is $\pi_7(F)=0$, where $F$ is the Yang-Mills
 field strength and $\pi_7$ is the projector onto two-forms that transform in the $7$ of $G_2$.  On a $\Spin(7)$ manifold, the equation is again
 $\pi_7(F)=0$, or alternatively $\star F+\Phi\wedge F=0$,
 where $\star $ is the Hodge star and $\Phi$ is the covariantly constant four-form, suitably normalized.
 These are the analogs of the hermitian Yang-Mills
 (or Donaldson-Uhlenbeck-Yau) equation that was employed in \cite{EW}.}  Replacing Type II superstrings by the heterotic string
 reduces the spacetime supersymmetry, and an argument based on holomorphy is not available.  Moreover, there is no useful $R$-symmetry.  So we expect
 that in this class of  heterotic string compactifications, $\alpha'$ corrections do spoil spacetime supersymmetry.

\section{$G_2$ Holonomy Manifolds}

\subsection{Preliminaries}\label{gtwom}

We consider Type II (either IIA or IIB) string theory in a spacetime of the form $\RR^{2,1}\times M$, where $M$ is
a compact seven-dimensional manifold. We will not turn on fluxes\footnote{For some analysis including $G$-flux at the classical level, see \cite{Kaste:2003zd}.}, and we wish to preserve a minimal amount of supersymmetry in three dimensions. The
condition for finding a supersymmetry generator leaving the vacuum invariant
imposes strong constraints on spacetime.\footnote{\label{indcon} In what follows, indices
$M,N,\dots$, $\m,\n,\dots$ and $a,b,\dots$ are tangent to
the ten-, three- and seven-dimensional spaces, respectively.}
For example, within supergravity it requires the existence of a covariantly constant spinor $\eta$ on $M$.
This implies that $M$ is Ricci-flat,
the three-form
\begin{equation}\label{spin1}
\vp_{abc} = i \eta^T \Gamma_{abc} \eta,
\end{equation}
and the dual four-form $\psi = \star \vp$ are covariantly constant, and the holonomy group of $M$ is $G_2$.
Such a spacetime solves the Einstein equations, and moreover
the string theory beta-functions vanish on such a solution, up to three-loop order.

However, the four-loop correction to the beta-function for the metric
does not vanish, in general,  for a $G_2$-holonomy space.  We would like to show that, as in the Calabi-Yau case,
we can always find a globally-defined $\alpha'$-dependent metric, which is close to the Ricci-flat metric, and which solves the equations
to all orders in $\a'$.

Locally, the moduli of the $G_2$ metric on $M$ are given by the cohomology class of the three-form $\vp$.  This cohomology
class takes values in a certain cone $\frak C\subset H^3(M,\Bbb{R})$, which in general is not well-understood.  The analysis of this
paper holds for any choice of the cohomology class of $\vp$, within the cone $\frak C$.

\subsection{Leading Order Correction}\label{lco}

Since in ten-dimensions the supersymmetry algebra only closes on-shell,
a correction to the equations of motion will, in general, lead to corrections
to the supersymmetry variations.
To parametrize these corrections, it is useful
to recall that in seven dimensions, the spinor representation is a real representation of dimension 8.
Suppose we have a nowhere-vanishing spinor field $\eta$ on $M$, which we can normalize so that $\eta^T\eta=1$.
Let $\Gamma_a, $ $a=1,\dots,7$ be the Dirac matrices, obeying $\{\Gamma_a,\Gamma_b\}=\delta_{ab}$.
One cannot choose the $\Gamma_a$ to be real, but one can choose them to be purely imaginary (so that the $SO(7)$
generators $\Gamma_{ab}=\frac{1}{2}[\Gamma_a,\Gamma_b]$ are real) and we will do so.
Any other real spinor $\psi$ can then be expanded in the basis
$\{\eta, i \Gamma_a \eta\}$,  $a=1,\dots, 7$. Explicitly
\begin{equation}
\psi =  \eta \left(\eta^T \psi\right) +\Gamma^a \eta \left( \eta^T \Gamma_a \psi\right) .
\end{equation}
In particular,
 by expanding in this basis,
we can encode any possible corrections to the transformation law of the ten-dimensional gravitino under
spacetime supersymmetry in terms of  two tensors $A_a$ and ${B_a}^b$:
\begin{equation} \label{di}
\d \psi _a = \zeta\otimes\lp D_a\eta+A_a\eta+iB_a^{\hph{a}b}\Gamma_b\eta\rp.
\end{equation}
Here $\zeta$ is a three-dimensional spinor, which we can take to be constant (in looking for Lorentz-invariant solutions in three dimensions).
To order $\a'^3$, the corrections to the supersymmetry transformation were  presented in ref. \cite{Lu:2003ze}:
\begin{equation}
\begin{split}
A_a & = 0 , \\
{B_a}^b & =  {c \o 2} \a'^3 \vp_{acd} \nabla^c Z^{db},
\end{split}
\end{equation}
where $Z^{ab}$ is the symmetric tensor built out of three Riemann tensors
\begin{equation}
Z^{ab}=\frac{1}{32 g }\e^{ac_1\cdots c_6}\e^{bd_1\cdots d_6}R_{c_1c_2d_1d_2}R_{c_3c_4d_3d_4}R_{c_5c_6d_5d_6},
\end{equation}
$c$ is a constant, and $g = \det g_{ab}$.

Given the $\a'$-corrected supersymmetry transformations, next we
construct a supersymmetric solution perturbatively in $\a'$. Henceforth we label quantities of the corrected
space with primes while unprimed quantities are unperturbed. So the $\a'$-corrected internal space is
denoted by $M'$.
To preserve supersymmetry we must find a globally-defined spinor $\eta'$ on $M'$ which equals $\eta$ to leading order in $\a'$
and obeys the appropriate $\a'$-deformed equation.
Since
$A,B$ are already $\O(\a'^3)$, the vanishing of the variation \C{di} to order $\a'^3$ becomes
\begin{equation} \label{dii}
D_a'\eta'+A_a\eta+iB_a^{\hph{a}b}\Gamma_b\eta=0.
\end{equation}
If we are given such an $\eta'$, even just defined locally, it is possible to construct a tensor
\begin{equation}\label{vpr}
\vp'_{abc} = i \eta'^T \Gamma_{abc}' \eta',
\end{equation}
(defining what is called a $G_2$-structure that may have torsion)
and an associated metric\footnote{Here $g'$ is defined in terms of $\vp'$ by the formula (\ref{metric1}).} $g'_{ab}$.
We also have
\begin{equation} \label{pspr}
\psi'_{abcd} = \eta'^T \Gamma_{abcd} ' \eta',
\end{equation}
or equivalently $\psi'=\star'\vp'$, where the Hodge star is constructed from $g'_{ab}$.

Eq. \C{dii} can be converted into conditions on $\vp'$ and $\psi'$ by
taking derivatives of eqns. \C{vpr} and \C{pspr}.
We define the four-form $\a$ and five-form $\b$
\begin{equation} \label{ai}
\begin{split}
\a & = d \vp', \\
\b & = d \ps',
\end{split}
\end{equation}
which are explicitly given by
\begin{equation}\label{aaii}
\begin{split}
\a_{abcd}  &  =8 A_{[a} \vp_{bcd]}-8B_{[a}^{\hph{[a}e}\psi_{bcd]e},\\
\b_{abcde} &  =10 A_{[a} \ps_{bcde]}-40B_{[ab}\vp_{cde]}.
\end{split}
\end{equation}
This establishes a connection between $A_a$ and ${B_a}^b$ and failure of $\phi'$ and $\psi'$ to be closed (and thus
to the torsion forms defined in
proposition 1 of ref. \cite{Bryant:2005mz}).

With a view to the generalization   beyond order $\a'^3$,  we have written the correction
in eqn.\ \C{dii} in terms of $A_a$ and ${B_a}^b$. So far these tensors could be anything.
However, from eqs.\ \C{ai} and \C{aaii} we see that
a necessary condition on $A_a$ and ${B_a}^b$
for the existence of a $G_2$-structure space $M'$ close to $M$ is that the forms
$\a$ and $\b$ are exact.  As we show next this condition is also sufficient.
To order $\a'^3$, we can use the explicit expressions above to show that $\a=d\chi$ and $\beta=d\xi$, with
\begin{equation}
\label{eq:LeadingOrderChi}
\begin{split}
\chi_{abc} & =-c\vp_{abc}Z+3c\vp_{[ab}^{\hph{[ab}d}Z_{c]d},\\
\xi_{abcd} & =-4c\psi_{[abc}^{\hph{[abc}e}Z_{d]e},
\end{split}
\end{equation}
where $Z=Z^a_{\hph{a}a}$. Thus $\a$ and $\b$ are exact to order $\a'^3$.

\subsubsection{Existence Of A Solution}\label{existence}

Our task is now
to find a globally-defined
$G_2$-structure $\vp'$ (and its associated metric $g'$ and four-form $\psi'$)
which is close to $\vp$, \ie
\begin{equation}
\vp' = \vp + \d \vp,
\end{equation}
and solves the equations
\begin{equation}
\begin{split}
d\vp' & =\a , \\
d\psi' & =\beta,
\end{split}
\end{equation}
for given exact forms $\a= d \chi $ and $\b= d\xi$.

We can satisfy the first equation by\footnote{\label{lobol} The possibility of adding to $\vp'$ a closed but
not exact three-form is not really interesting here, because this could be absorbed in shifting the cohomology
class of the starting three-form $\vp$.  As remarked at the end of section \ref{gtwom}, this cohomology class is arbitrary (within a certain cone).
So the form given in (\ref{biv}) is essentially the most general.}
\begin{equation}\label{biv}
\vp'=\vp+\chi+db,
\end{equation}
for some two-form $b$. What about the second one?
To leading order in fluctuations the dual four-form $\psi' = \star' \vp'$ satisfies
\begin{equation}
\label{bi}
d\psi'=d\star\lp\frac{4}{3}\pi_1+\pi_7-\pi_{27}\rp\lp\chi+db\rp.
\end{equation}
 Here $\pi_1,\pi_7,\pi_{27}$ are the
projections of three-forms onto $\wedge ^3_1, \wedge ^3_7$ and $\wedge ^3_{27}$ respectively (these are the subspaces of $\wedge ^3$ that transform in the indicated
representations of $G_2$; see the appendix
for concrete expressions).    Deformations of $G_2$-structures
have been studied in the literature (see, for example,
\cite{kari,Grigorian:2008tc,Grigorian:2009ge}), and the tools developed in that context are useful in what follows.
  All we need to know to  derive eqn.\ \C{bi} is described in the appendix. Besides the explicit $\d \vp$ appearing in $\vp'$, there is
also implicit $\d \vp$ dependence in $\star'$ since the metric is a functional of $\vp'$. Taking
this fact into account one easily derives eqn.\ \C{bi}.

We get from (\ref{bi}) an equation for $b$, which can be written as
\begin{equation}\label{zi}
d^\dagger\left(\pi_{27}-\pi_7-\frac{4}{3}\pi_1\right)d b = d^\dagger\rho,~~~\rho=-\star\xi-\left(\pi_{27}-\pi_7-\frac{4}{3}\pi_1\right)\chi.\end{equation}
Here $d^\dagger=-\star \negthinspace d\star$ in acting on three-forms in seven dimensions.

To proceed further, we decompose $b$ into irreducible representations of $G_2$.
The space of two-forms decomposes as $\wedge ^2 = \wedge ^2_7 \oplus \wedge ^2_{14}$, where 7 and 14 are the representations of $G_2$
of the indicated dimension. Actually,
the component of $b$ in $\wedge ^2_7$ does not contribute to the right hand side of eqn.\ \C{bi},
as is shown in eqn.\ \C{eq:7DropsOut} of the appendix.
Hence we can assume that $b\in \wedge ^2_{14}$.
  The reason that the part of $b$ in $\wedge ^2_7$ does not contribute to the equation for unbroken supersymmetry is that for
 $b \in \wedge ^2_7$, $d b$ is the  change in $\vp$ generated by an infinitesimal diffeomorphism.   Indeed,
a general section of $\wedge ^2_7$ is  $b_{ab}={\vp_{ab}}^c v_c$ for some vector field $v^c$, and  the corresponding change in the metric is
$\d g_{ab} = 2 \nabla_{(a} v_{b)}$, which is the first-order change in the metric generated by  $v$.

It is also true that $d^\dagger\rho\in \wedge ^2_{14}$.  This elementary but somewhat tricky fact is explained in the appendix (see
also eqn.\ (\ref{rolx}) for another point of view).

We want to show that equation (\ref{zi}) has a solution for $b$.  To do this, first let $\Delta=d^\dagger d+d d^\dagger$ be the Hodge-de Rham
Laplacian, and observe that it is possible to solve the equations
\begin{equation}\label{pix}\Delta b=d^\dagger\rho, ~~d^\dagger b=0. \end{equation}
Indeed, standard Hodge theory says that a two-form $b$ obeying these equations always exists for any three-form
$\rho$ on any compact
manifold, since $d^\dagger\rho$
is orthogonal to the harmonic two-forms.  Moreover, on a $G_2$ manifold, $\Delta$ preserves the decomposition
$\wedge ^2=\wedge ^2_{14}\oplus \wedge ^2_7$, so a solution exists with $b\in\wedge ^2_{14}$.
We can write the first equation in (\ref{pix}) as
\begin{equation}\label{modz} d^\dagger\left(\pi_{27}+\pi_7+\pi_1\right)d b =d^\dagger\rho,\end{equation}
since $\wedge ^3=\wedge ^3_{27}\oplus \wedge ^3_7\oplus \wedge ^3_1$.  But actually, the $\pi_1$ and $\pi_7$ terms do not contribute in
eqn.\ (\ref{modz}).  $\pi_1$ does not contribute because for any $b\in\wedge ^2_{14}$, $\pi_1db=0$; this is because the representation 1
of $G_2$ does not appear in the decomposition of $7\otimes 14$, so it does not appear in the first derivatives of $b$.  Also, if $d^\dagger b=0$,
one has $\pi_7(db)=0$; this is because the 7 of $G_2$ appears only once in the decomposition of $7\otimes 14$, so there is essentially
only one way to form linear combinations of the first derivatives of $b$ transforming as the 7 of $G_2$, and hence $\pi_7(db)$ is linear in
$d^\dagger b\in \wedge ^1=\wedge ^1_7$.  Since the $\pi_1 $ and $\pi_7$ terms do not contribute, eqn.\ (\ref{modz}) is equivalent to eqn.\ (\ref{zi}),
which we wished to solve.

\subsection{All Orders In $\a'$}\label{allex}

Next we describe the generalization to arbitrary orders in $\a'$.
We are given supersymmetry transformations
\begin{equation}
\label{eq:GeneralG2GravitinoVar}
\d \psi_a = \zeta \otimes \left( D'_a \eta' + A_a \eta' + i {B_a}^b \Gamma_b' \eta'\right).
\end{equation}
Here $A,B$ are functionals of $\vp'$:
\begin{equation}
A_a=A_a[\vp'], \qquad {B_a}^b={B_a}^b [\vp'].
\end{equation}
This time we include all orders in the derivative expansion. So
\begin{equation}\label{gi}
B[\vp'] = \sum_{n=3}^\infty B_n [\vp'],
\end{equation}
where each $B_n [\vp']$ is a local covariant expression (or functional) constructed out of\footnote{In the $\alpha'$ or $1/\br$ expansion, the effective action and (therefore) the supersymmetry
transformations are constructed in terms of $g'$ without reference to $\vp'$, so actually  $B$ can be constructed
from the metric $g'$ and its derivatives, without reference to $\vp'$.
(Note that $g'$ can be expressed in terms of $\vp'$ via (\ref{metric1}), but not the other way around.
In the tangent space at any
point, $g'$ has $SO(7)$ symmetry and $\vp'$ reduces the symmetry to $G_2$.)
To leading order, the expression for $B$ in terms of the metric was given in
\cite{Lu:2003ze}.} $\vp'$
and its associated metric $g'$,
Riemann tensor $R'$, and covariant derivatives $\nabla'$, and where
$B_n[\vp']$
contains $2n+1$ explicit derivatives (with each Riemann tensor counting two and each covariant derivative counting one).
There is a similar
expansion for $A_a$.

We can then construct $\vp'$ and its dual $\psi' = \star' \vp'$
\begin{equation}
\begin{split}
\vp' & = \eta'^T \Gamma_{abc}' \eta', \\
\psi' & = \eta'^T \Gamma_{abcd}' \eta',
\end{split}
\end{equation}
and use the condition for unbroken supersymmetry to compute
\begin{equation}\label{giv}
\begin{split}
d \vp' & = \a[\vp'], \\
d\psi' & = \b[\vp'].
\end{split}
\end{equation}
Explicitly
\begin{equation}\label{aii}
\begin{split}
\a_{abcd}  &  =8 A_{[a} \vp'_{bcd]}-8B_{[a}^{\hph{[a}e}\psi_{bcd]e}',\\
\b_{abcde} &  =10 A_{[a} \ps_{bcde]}'-40B_{[ab}\vp_{cde]}'.
\end{split}
\end{equation}

Next we wish to construct the supersymmetric background perturbatively in $\a'$. We denote the order $n$ term of  $\vp'$ by $\vp_n$.
We proceed by induction.
In our previous analysis, we constructed $\vp_3$ explicitly. We assume the $G_2$-structure
is known up to order $n-1$ in $\a'$. We denote this $G_2$-structure by
\begin{equation}
\tilde \vp = \vp + \sum_{k=3}^{n-1} \vp_k,
\end{equation}
and we construct the order $n$ contribution to $\vp'$ as a small perturbation around $\tilde \vp$
\begin{equation}
\vp' = \tilde \vp + \d \vp ,
\end{equation}
where $\d \vp = \vp_n$. This deformation of $\tilde \vp$ leads to a deformation of
$\ps'$
\begin{equation}
\ps' = \tilde \ps + \d \ps,
\end{equation}
where $\tilde \ps = \tilde \star \tilde \vp$ and
\begin{equation}
\d \ps = \tilde \star \left( {4 \o 3} \tilde \pi_1 + \tilde \pi_7 - \tilde \pi_{27} \right) \d \vp,
\end{equation}
where $\tilde \star$ and $\tilde \pi$ are the Hodge dual and projection operators with respect to the $G_2$-structure $\tilde \vp$. We will label
the term of  order $n$ in $\a'$ of any quantity by $\mid_n$ and from the above we see that the order $n$ of $\ps'$ is
\begin{equation} \label{eqnai}
\psi'\mid_n = \tilde \star \tilde \vp  \mid_n +  \star \left( {4 \o 3} \pi_1 +  \pi_7 - \pi_{27} \right)\vp_n ,
\end{equation}
Note that $\tilde \vp$ includes terms only up to order $n-1$ but $\tilde \star$ and $\tilde \psi$ can, in general, receive contributions
of any order since they are non-linear functionals of $\tilde \vp$.

Let us suppose that the dependence of $A_a$ and ${B_a}^b$ on $\vp'$ is such that  order by order in $\a'$ the forms $\a$ and $\b$ are exact.  (In section \ref{coco}, we interpret this statement in terms of  effective
field theory.)  Then there exist globally-defined
$\chi_n$ and $\xi_n$ with
\begin{equation}\label{exactai}
\begin{split}
\a[\vp']\mid_n & = d \chi_n, \\
\b[\vp']\mid_n & = d \xi_n.
\end{split}
\end{equation}
Note that since $\a$ and $\b$ are already order $(\a')^3$ explicitly, we can view $\chi_n$ and $\xi_n$ as being functionals of $\tilde\vp$; they do not
depend on $\vp_n$.

By setting
\begin{equation}\label{exactaii}
\vp_n = \chi_n + db_n,
\end{equation}
we solve the part of the first equation in eqn.\ \C{giv} that is of  the order $n$ in $\a'$.
The second equation in \C{giv}
turns into a partial differential equation for $b_n$. Indeed, we take the exterior derivative of \C{eqnai} and set $d \psi'\mid_n = d \xi_n$, so
\begin{equation}
d \xi_n = d (\tilde \star \tilde \vp) \mid_n + d \left[ \star \left( {4 \o 3} \pi_1 +  \pi_7 - \pi_{27} \right)\right](\chi_n + d b_n).
\end{equation}
This can be recast in the form
\begin{equation} \label{gv}
\D b_n = d^\dagger \r_n,\quad d^\dagger b_n =0,
\end{equation}
with
\begin{equation}
\star \r_n =\xi_n  -
\star\left({4\o 3} \pi_1 + \pi_7 -\pi_{27}  \right) \chi_n -\tilde \star \tilde \vp \mid_n.
\end{equation}
In complete analogy to the leading order case, any piece $b_n \in \wedge ^2_7$ drops out of eqn.\ \C{gv}.
Taking $b_n \in \wedge ^2_{14}$, we obtain eqn.\ \C{gv} after choosing the gauge $d^\dagger b_n=0$.
The source is co-exact, and satisfies $\pi_7(d^\dagger\rho_n)=0$ (this is shown in Appendix \ref{app:G2Torsion}), and the remaining steps
follow word by word the reasoning we used to leading order in $\a'$.

\subsection{Interpretation In Three- or Four-Dimensional Field Theory}\label{coco}

\def\S{{\mathcal S}}
\def\bar{\overline}

\subsubsection{Superfields}\label{super}
We aim here to interpret the above results in the effective field theory  that arises by compactification
of Type IIA or Type IIB superstring theory on a $G_2$ manifold $M$. This is a theory with $\N=2$ supersymmetry (four supercharges)
in three dimensions.  In studying this theory, we will go beyond low energy effective field theory and include Kaluza-Klein harmonics in
a way that preserves three-dimensional supersymmetry.  We should warn the reader that
more work is needed to fully justify the way we do this.  Our analysis
is somewhat speculative.

We will make the analysis for Type IIA and leave the analogous story for Type IIB for the future.
Rather than the $\alpha'$ expansion of Type IIA on $\R^3\times M$, we can in a similar way study the $1/\br$
expansion of M-theory on $\R^4\times M$ (here $\br$ is the radius of $M$).    It is not completely trivial that the analysis is the same for Type IIA and for M-theory, because
in general $\N=2$ theories in three dimensions, there can be supersymmetric interactions
(Chern-Simons couplings of vector multiplets, for instance) that do not arise by classical dimensional reduction from four dimensions.
If these were important, the Type IIA and M-theory analyses would be essentially different.  However, the supersymmetric interactions
 that will be relevant are the most basic ones related to superpotentials and Kahler potentials for chiral superfields,
 and these are possible in four dimensions.

Hence at the general level of the following discussion, the constraints on the $\alpha'$ expansion of Type IIA are the same as those on the
$1/\br$ expansion of M-theory and the structure we will find applies to each.   Our analysis is limited to finite orders in $\alpha'$ or $1/\br$
because we assume locality along $M$, which nonperturbatively in $\alpha'$ is violated by worldsheet instantons (Type IIA), and nonperturbatively
in $1/\br$ is violated by M2-brane instantons (M-theory).  What happens nonperturbatively in $1/\br$ is quite different from what happens
nonperturbatively in $\alpha'$.  In M-theory, the M2-brane instantons violate a certain shift symmetry (adding a harmonic form to the $C$-field)
and generate a spacetime superpotential \cite{HM} that destabilizes the $\R^4\times M$ compactification.  In Type IIA, the worldsheet instantons respect
the relevant shift symmetry\footnote{They violate a different shift symmetry involving the $B$-field of the NS sector.}
 and a simple argument given in the introduction shows that, to all finite orders in the string coupling constant
$g_{st}$, the $\R^3\times M$ compactification remains supersymmetric.\footnote{Nonperturbatively in $g_{st}$, this stability is lost because of D2-brane
instantons, which are analogs of M2-brane instantons in M-theory.}  In particular, setting $g_{st}=0$, one expects an exact superconformal
field theory describing $G_2$ compactification.  The more detailed analysis we give here, which aims to explain what we have
found in sections \ref{gtwom}-\ref{allex}, assumes locality along $M$ and so is valid only to all finite orders in $\alpha'$.

Exploiting three- or four-dimensional supersymmetry as well as locality along $M$,
 we will try to describe Type IIA or M-theory compactification on $M$ in terms of three- or four-dimensional
superfields that are also functions, or forms, along $M$.  This part of the analysis will be more
transparent in  the M-theory language.  Once we have identified the relevant superfields, we will phrase our discussion in terms of the
$\alpha'$ (rather than $1/\br$) expansion.

The bosonic fields of eleven-dimensional supergravity consist of the metric tensor\footnote{See footnote \ref{indcon} for index conventions.}
 $g_{MN}$ and the three-form field $C_{MNP}$.  In compactification on $M$, these fields will give the propagating bosonic modes of four-dimensional supermultiplets.  We can see
what these supermultiplets must be as follows:

(1)  The part $g_{\mu\nu}(x;y)$ of the metric tensor (here $x$ is a coordinate along $\R^4$ and $y$ along $M$) gives fields of spin 2 or 0
on $\R^4$ that are scalar functions on $M$.  Clearly, the Kaluza-Klein expansion of M-theory on $\R^4\times M$ contains massive spin 2 supermultiplets. We will not try to understand these multiplets here, though this will have to be part of a full understanding.

(2) There are two sources of spin 1 fields along $\R^4$ that will be the bosonic parts of vector multiplets.  From the $C$-field, we get
the components $C_{\mu ab}(x;y)$, which we interpret as the bosonic part of a vector multiplet $V_{ab}$ that is a two-form along $M$
(and a vector multiplet in $\R^4$).  Similarly, the components $g_{\mu }{}^a$ of the eleven-dimensional metric give vector multiplets $\h V^a$
that comprise a vector field  along $M$.

(3) Taking advantage of the fact that M-theory on $\R^4\times M$ is invariant under a reflection of $\R^4$ combined with a sign change of $C$,
spin zero fields in this theory can be classified as scalars or pseudoscalars.  Pseudoscalars come from the part $C_{abc}(x;y)$ of the $C$-field,
which gives us a pseudoscalar field on $\R^4$ that is a three-form on $M$.  Scalars come from the part $g_{ab}$ of the metric and also from the
part $C_{\mu\nu a}$ of the $C$-field (here one must recall that a two-form on $\R^4$, such as $C_{\mu\nu a}$, is dual to a field of spin 0).  We expect all these fields to
combine to the propagating modes of a field $\h\Cc_{abc}$ that will be a chiral superfield on $\R^4$ and a three-form on $M$.  The bottom component
of $\h\Cc_{abc}$ is a complex field of spin 0 that we will call $\Cc_{abc}$.  We expand $\Cc_{abc}(x;y)=\h\phi_{abc}(x;y)+iC_{abc}(x;y)$, where the
imaginary part is the pseudoscalar field $C_{abc}$, and the real part $\h\phi_{abc}$ is constructed from $g_{ab}$ and (the dual of) $C_{\mu\nu a}$.
Concerning this last point, we note that in expansion around a metric of $G_2$ holonomy, a three-form $C_{abc}$ transforms as $1\oplus 7\oplus 27$,
while a perturbation in the metric $g_{ab}$ transforms as $1\oplus 27$, and the dual of\footnote{The dual of $C_{\mu\nu a}$ is even under
$C\to -C$ accompanied by a reflection of one coordinate in $\R^4$, so $\hat\phi_{abc}$ is even under this symmetry while $C_{abc}$ is odd.
Hence the Kahler potential that we introduce later is invariant under $\hat\phi\to\hat\phi$, $C\to -C$.}
 $C_{\mu\nu a}$ transforms as 7.  So the pieces are
there for the scalar and pseudoscalar fields to combine properly into the complex three-form  $\Cc_{abc}$.  The reason that we denote
the real part of $\Cc_{abc}$ as $\h\phi$ is that in the classical limit ($1/\br\to 0$ or $\alpha'\to 0$), expanding around a metric of $G_2$ holonomy,
$\h\phi$ will coincide with the covariantly constant three-form $\phi$ associated to the $G_2$ metric.   Thus $\h\phi$ will be an $\alpha'$-corrected
version of $\phi$.
$\h\phi$ will be the analog in our present analysis of the $\alpha'$-corrected three-form that was called $\phi'$ in section \ref{lco}.  The relation
between $\phi'$ and $\h\phi$ will be the subject of section \ref{relex}.

We can use the chiral multiplets and vector multiplets that have just been introduced to describe M-theory on
$\R^4\times M$, to all finite orders in $1/\br$, in a way that exhibits supersymmetry
(and locality) along $\R^4$ and locality along $M$.  The same set of superfields can be used to similarly describe Type IIA on $\R^3\times M$.
Rather than always repeating ourselves, we formulate the following statements in terms of Type IIA.
In what follows, the goal will be to describe possible supersymmetric vacua.  In such a vacuum, the fields are independent of $x$.
So we can drop the dependence of the fields on $x$ (that is, on $\R^3$ or $\R^4$) and concentrate on the dependence on $y$.

\subsubsection{Conditions For Unbroken Supersymmetry}\label{consym}

In general, in a globally supersymmetric four-dimensional theory of chiral multiplets $\Phi$ and vector multiplets $V_\zeta$,
or a three-dimensional theory of such multiplets with no
Chern-Simons couplings for the vector multiplets,\footnote{In the present context, the symmetry of Type IIA on $\R^3\times M$ under a reflection of $\R^3$ (together
with a reversal of orientation of the string worldsheet) ensures that Chern-Simons couplings are not possible for the vector multiplets $V_{ab}$
and $\h V_a$.}  the condition for unbroken supersymmetry is $\delta W=0=D_{\zeta}$, where $\delta W$ is the variation of the spacetime superpotential
$W(\Phi)$,
and $D_\zeta$ are the auxiliary fields in the vector multiplets $V_\zeta$.
After coupling to supergravity, to get unbroken supersymmetry in Minkowski spacetime, one additionally needs $W=0$.

 In the present context, we can easily make explicit the conditions
$W=\delta W=0$.  $W$ will have to be a holomorphic function of the three-form field $\Cc$.  It must be constructed without use of a metric on $M$
(since the metric is a function of  the real part of $\Cc$ and thus is a non-holomorphic function of $\Cc$).  So up to a constant multiple, the superpotential
must be
\begin{equation}\label{zomo} W=\int_M\Cc\wedge d \Cc. \end{equation}
The condition $\delta W=0$ is thus simply
\begin{equation}\label{womo} d\Cc=0.  \end{equation}
In more detail, this is
\begin{equation}\label{omom}G=0,\end{equation}
where $G=d C$ is the field strength of the three-form field $C$ (or more precisely the part of this field strength
that is a four-form along $M$), and
\begin{equation}\label{tomo} d\h\phi=0. \end{equation}
Thus in contrast to section \ref{lco} where the $\alpha'$-corrected three-form $\phi'$ did not obey $d\phi'=0$, here there will
be no $\alpha'$ correction to the statement $d\h\phi=0$.    The further condition $W=0$ is automatic if $\Cc$ is a globally-defined three-form (for then
$d\Cc=0$ implies that $\int \Cc\wedge d\Cc=0$);
since $\hat\phi$ is certainly globally-defined, this is true if the $C$-field is topologically trivial.

To learn more,  we will have to impose the second condition for unbroken supersymmetry,
which is the vanishing of the auxiliary fields $D_\zeta$.
As explained above, the theory of interest has
two kinds of vector multiplets, namely a two-form $V_{ab}$ and a vector field $\h V^a$.
The symmetry gauged by $\h V^a$ is the group of diffeomorphisms of $M$.  On the other hand,
the two-form $V_{ab}$ gauges the group of $C$-field gauge transformations $C\to C+d\Lambda$, $\h\phi\to \h\phi$,
where $\Lambda$ is
an arbitrary two-form on $M$.  Thus the transformation is
\begin{equation}\label{zobz}\delta\Cc=i d\Lambda. \end{equation}
Clearly, $V_{ab}$ is odd under $C\to -C$, while $\h V^a$ is even. It turns out that the corresponding auxiliary field $D_V$ and $D_{\h V}$ transform
oppositely to $V$ and $\h V$ (this is because the Kahler form that we analyze below is odd under $C\to -C$), so $D_V$ is even and $D_{\h V}$ is
odd.  $D_V$ and $D_{\h V}$ are gauge-invariant local functionals of $C$, so they are really depend on $C$ only through $G=dC$.  Eqn. (\ref{omom}) tells us
that  $G=0$
in a supersymmetric vacuum, and once we set $G=0$, $D_{\h V}$ will automatically vanish, since it is odd in $G$.  So the additional condition for unbroken supersymmetry
that we are looking for will be $D_V=0$.

In general, consider any theory with four supercharges with chiral multiplets $\h\Cc=\Cc+\dots $ that parametrize a Kahler manifold
$\X$, and with vector multiplets $V_\zeta$ generating a group $\G$ of symmetries of $\X$.
On-shell, the corresponding auxiliary fields $D_\zeta$ are functions $D_\zeta(\Cc,\bar\Cc)$ on $\X$ and together
these functions comprise the
``moment map'' for the action of $\G$ on $\X$.  In general, these functions transform in the representation of $\G$ that is dual
to the adjoint representation.
In the present context, we take $\G$ to be the group of $C$-field gauge transformations, so 
a generator of  $\G$ is a two-form $\Lambda$, as in eqn.\ (\ref{zobz}).  Hence, the auxiliary field 
will be a five-form $D(\Cc,\bar\Cc)$ which, order by order in the $\alpha'$ expansion, will be constructed locally from $\Cc$, $\bar \Cc$, and their derivatives.

But actually, the group $\G$ does not act faithfully on $C$: the $C$-field gauge transformation generated by a {\it closed} two-form $\Lambda$ is trivial.
Hence $D(\Cc,\bar\Cc)$ takes values in the  dual to the space of two-forms modulo closed ones.  In seven dimensions, the dual of two-forms mod closed ones  is the space of {\it exact} five-forms,
and therefore $D(\Cc,\bar\Cc)$ will be exact, as we find explicitly below.

In our problem, the Kahler manifold $\X$ parametrized by the $\Cc$'s is simply the space of complex-valued three-forms on $M$.
We must discuss the Kahler metric on $\X$, since in general the moment map depends on the Kahler metric.
The Kahler metric $d s^2$ is determined in the usual fashion by a Kahler potential $K(\Cc,\bar\Cc)$:
\begin{equation}\label{mmi}d s^2 =\int_{M\times M}\d \Cc(y)\otimes \d \bar \Cc(y')\frac{\delta^2 K}{\d \Cc(y)\d \bar\Cc(y)}. \end{equation}
(We will write $\d$ for a variation on the infinite-dimensional space $\X$, and $d$ for the exterior derivative on $M$.)
Though we have written the metric as an integral over $M\times M$, this integral actually collapses, order by order in
the $\alpha'$ expansion, to an integral over a single copy of
$M$.  The reason for this is that the functional $K$ is local order by order in $\alpha'$
(that is, it is the integral over $M$ of a local function of $C$, $\hp$, and their derivatives up to a finite order), so that the second variation
$\d^2 K/\d\Cc(y)\d\bar\Cc(y')$ is a sum of terms proportional to $\delta(y,y')$ and its derivatives.  This ensures that the integral
on the right hand side of (\ref{mmi}) collapses, order by order, to a local integral over $M$.

Type IIA superstring theory has
 a symmetry $(-1)^{F_L}$ that ensures that $K$ is an even function of $C$.  Also, the Kahler metric of $\X$ does not depend on the
 orientation\footnote{As explained in footnote \ref{welf} of the appendix, a $G_2$ manifold does not have
 a preferred orientation.} of $M$, since Type IIA superstring theory on $\R^3\times M$ is invariant under simultaneous reversal
 of $\R^3$ and $M$.  The Kahler potential $K$ of $\X$ is not uniquely determined, since one is free to make a Kahler transformation
 $K\to K+f+\bar f$, where $f$ is a holomorphic function on $\X$.  But since the Kahler metric of $\X$
 is gauge-invariant, $K$ can be chosen
 to be gauge-invariant and thus to depend on $C$ only via $G=dC$.  (The fact that $K$ does not depend on the orientation of $M$
 excludes a term $\int_M C\wedge d C$ in $K$.)

The fact that $K$ depends on $C$
only via $G=d C$ will ensure that the $C$-dependent terms in $K$ do not contribute to the functions $D(\Cc,\bar \Cc)$.
To explain this, we will expand $K$ in powers of $G$, keeping the quadratic term
 (there is no linear term as $K$ is even in $C$).   It will be clear from the derivation that terms in $K$
 that are higher than quadratic in $G$ will  not contribute to $D(\Cc,\bar\Cc)$ at $G=0$,
 and actually we will see that also the quadratic terms do not contribute.  Thus we write
\begin{equation}\label{molb}K= 4K_0(\hp)+\int_{M\times M}G(y) \wedge G(y') \wedge K_1(y,y')+\dots.\end{equation}
(The factor of 4 is for later convenience.)
Here $K_1(y,y')$ is a three-form in each variable.  As in (\ref{mmi}), we write the second term here as an integral over $M\times M$,
but order by order in $\alpha'$, it actually collapses to an integral over a single copy of $M$, since $K_1(y,y')$ is proportional
to $\delta(y,y')$ and its derivatives up to a finite order.
The Kahler metric derived from (\ref{molb}) is
\begin{equation}\label{polb}d s^2=\int_{M\times M}\d \Cc(y) \otimes \d\bar\Cc(y')\frac{\delta^2 K_0}{\delta \h\phi(y)\delta\h\phi(y')}+
\frac{1}{2}\int_{M\times M}d \delta
\Cc(y) \wedge d\delta \bar\Cc(y')\wedge K_1(y,y')+\O(G) .\end{equation}
We omit terms proportional to $G$ as we will be setting $G=0$.
The corresponding Kahler form is
\begin{equation}\label{zolb} \omega=\int_{M\times M}\d \h\phi(y)\wedge \d C(y')\frac{\delta^2 K_0}{\delta \h\phi(y)\d \h\phi(y')}
+\frac{1}{2}\int_{M\times M}d\d\h\phi(y)\wedge d\d C(y') \wedge K_1(y,y').\end{equation}

In general, given a vector field $V^\zeta$ acting on a Kahler manifold $\X$ with Kahler form $\omega$, the corresponding $D$-auxiliary field
$D_\zeta$ is characterized (up to an additive constant which is known as the Fayet-Ilioupoulos $D$-term) by the relation $\delta D_\zeta=\iota_{V^\zeta}\omega$, where $\iota_{V^\zeta}$ is the operation of contraction with respect to $V^\zeta$.  To implement this in the present context,
we let $V^\Lambda$ be the vector field that generates the $C$-field gauge transformation $\delta C=d\Lambda$, for some two-form $\Lambda$.
The contraction $\iota_{V^\Lambda}\omega$ is evaluated by replacing $\delta C$ in the formula for $\omega$ by $-d\Lambda$
(there is a minus sign because $\iota_{V^\Lambda}$ anticommutes with $\d\h\phi$).  When we do this, the $K_1$ term does not contribute
because $d\d C(y')$ is replaced by $d^2 \Lambda(y')=0$.  So
\begin{equation}\label{olb} \iota_{V^\Lambda}\omega=-\int_{M\times M}\d\h\phi(y) \wedge d \Lambda(y')\wedge \frac{\delta^2 K_0}{\delta \h\phi(y)\delta
\h \phi(y')}. \end{equation}
We are now supposed to set this equal to $\delta D_\Lambda$.  Clearly  the desired relation $\delta D_\Lambda=\iota_{V^\Lambda}\omega$
is satisfied by\footnote{Diffeomorphism invariance does not allow us
to add an FI term to this formula.  An FI term would be a contribution to $D_\Lambda$ that is constant, that is independent of $\h\phi$.
But there is no diffeomorphism-invariant way to construct a five-form contribution to $D_\Lambda$ if it is not allowed to depend on the only
field in the problem, namely $\h\phi$.}
\begin{equation}\label{mollb} D_\Lambda= \int_M \Lambda \wedge d \frac{\delta K_0}{\delta\h\phi}. \end{equation}
The condition that $D_\Lambda=0$  for all $\Lambda$ is
that $D=0$, where
\begin{equation}\label{tolb} D=d \h\psi, \end{equation}
with
\begin{equation}\label{opolb}\h\psi=\frac{\delta K_0}{\delta\h\phi}. \end{equation}
Clearly, the five-form $D$ is always exact, even when it does not vanish.
We expect that $D$ corresponds to the exact five-form  $\beta$ that was found in sections \ref{lco} and \ref{allex}.  The conditions for unbroken supersymmetry
are $\delta W=D=0$, or in other words $d\h\phi=d\h\psi=0$.

In the classical limit $\alpha'\to 0$, $K_0$ is \cite{HM,GP,BW} a multiple of $\log V(M)$, where\footnote{See also \cite{Hitchin} for properties of this
volume functional.}  $V(M)$ is the volume of $M$ (computed using the metric
(\ref{metric1}), where in the classical limit, we need not distinguish $\h\phi$ from $\phi$).  With $K_0$ a constant multiple of $\log V(M)$,
$\h\psi$ as defined in (\ref{opolb}) is a multiple of $\psi=\star\phi$, where $\star$ is the Hodge star defined using the metric (\ref{metric1}).
So the conditions $\delta W=D=0$ give the expected results $0=d\phi=d\star\phi$, which characterize $G_2$ holonomy at the classical level.
In our framework, $K_0$ and therefore the function $\h\psi(\h\phi)$ will receive corrections order by order in $\alpha'$, and these give the
corrections to $G_2$ holonomy.  As for what metric should be used to describe $M$, taking into account the $\alpha'$ corrections, this
question does not have a unique answer.  One can simply use eqn.\ (\ref{metric1}) to define a metric $g$ on $M$, using $\h\phi$ instead of $\phi$.
With $\alpha'$ corrections included, this metric will not have $G_2$ holonomy, since although there is a closed four-form $\h\psi$,
it is not simply $\star_{\h\phi}\h\phi$ (where $\star_{\h\phi}$ is defined using the metric $g$).  Since this metric does not have
$G_2$ holonomy, it is not really distinguished.  One can just as well modify eqn.\ (\ref{metric1}) by adding $\alpha'$ corrections in the relation
between $g$ and $\h\phi$.

\subsubsection{Diffeomorphism Invariance And Expansion In Powers Of $\alpha'$}\label{diffex}

In this analysis, we have made no use of the second set of vector multiplets $\h V^a$, which gauge the diffeomorphisms of $M$.  Their $D$ auxiliary
fields trivially vanish once we set $G=0$, so the vanishing of these fields gives no additional constraint.  However, the existence of this gauge
symmetry means that the functional $K_0(\h\phi)$ is diffeomorphism-invariant.  Let us determine the implications of this.  Let $v^a$ be a vector
field on $M$ that generates an infinitesimal diffeomorphism.  The transformation of a general three-form $\h\phi$ generated by this vector field is
$\delta \h\phi=(\iota_v d+d \iota_v)\h\phi$, where $\iota_v$ is the operation of contraction with $v$.  If we restrict ourselves to the case that $d\h\phi=0$,
then
\begin{equation}\label{omox} \delta\h\phi=d\iota_v\h\phi. \end{equation}
Diffeomorphism invariance of $K_0$ means that (if $d\h\phi=0$) $K_0$ is invariant under this transformation,
so
\begin{equation}\label{polx}0=\int_M d\iota_v\h\phi\wedge\frac{\delta K_0}{\delta\phi}=\int_Md\iota_v\h\phi\wedge\h\psi.\end{equation}
After integrating by parts, and recalling that $d\h\psi$ is the exact five-form $D$, we learn that for any vector field $v$ on $M$,
\begin{equation}\label{wolx}0=\int_W \iota_v\h\phi\wedge D. \end{equation}
Now recall that any $\h\phi$ that obeys some mild inequalities such that the formula (\ref{metric1}) defines a Riemannian metric $g$
gives a reduction of the structure group of the tangent bundle of $M$ to $G_2$; this is known as a $G_2$-structure on $M$.   In particular, if $\h\phi$
arises in $\alpha'$ perturbation theory
starting with the covariantly constant three-form $\phi$ of a classical metric of $G_2$ holonomy, then certainly the relevant inequalities are obeyed
and $\h\phi$ does define a $G_2$-structure.  This means that it determines a decomposition of the space of two-forms on $M$ as $\wedge^2=
\wedge^2_7\oplus \wedge^2_{14}$, with corresponding projection operators $\pi_7$ and $\pi_{14}$.
Since $\iota_v\h\phi$ is an arbitrary section of $\wedge^2_7$, eqn.\ (\ref{wolx}) is equivalent to the identity
  \begin{equation}\label{rolx}\pi_7(D)=0 \end{equation}
  which holds whenver $d\h\phi=0$.

Let us now make a few remarks about solving the equations $d\h\phi=d\h\psi=0$ in an expansion in powers of $\alpha'$.  $K_0$ has an expansion
in powers of $\alpha'$ with the leading term being the classical expression $K_{0,cl}$ and the first correction being of order $(\alpha')^3$:
\begin{equation}\label{zelmo}K_0=K_{0,cl}+\sum_{n=3}^\infty (\alpha')^n K_{0,n}. \end{equation}
Correspondingly we expand $\h\phi$ in a series in $\alpha'$.  The classical term is the covariantly constant three-form $\phi$ of a classical metric
of $G_2$ holonomy.  The corrections must preserve the fact that $d\h\phi=0$, and (for the same reason as in footnote \ref{lobol}) we may as well
assume that the corrections do not change the cohomology class of $\h\phi$.  Therefore we assume that the series takes the form
\begin{equation}\label{welmo}\h\phi=\phi+\sum_{n=3}^\infty (\alpha')^n db_n, \end{equation}
with two-forms $b_n$.   In this expansion, we can assume that $b_n\in \wedge^2_{14}$ for the same reason as in section \ref{existence}.  Indeed, we are only interested
in the solution of the equations $d\h\phi=d\h\psi=0$ up to diffeomorphism.  In each order in the expansion, an infinitesimal diffeomorphism
(with a generator of order $(\alpha')^n$) can shift $\pi_7(b_n)$ in an arbitrary fashion, and therefore in solving the equations we can assume
that $\pi_7(b_n)=0$.
The equation $d\h\psi=0$, where $\h\psi=\delta K_0/\delta\h\phi$ and $K_0$ has the expansion (\ref{zelmo}), can then be expanded
in powers of $\alpha'$.  In order $(\alpha')^n$, with $n\geq 3$, we get a linear equation for $b_n$ that has the structure explored in sections
\ref{existence}-\ref{allex}.  This equation has an essentially unique solution  for the same reasons as described there.

\subsubsection{The Relation Between The Two Expansions}\label{relex}

The expansion that we have just described is very similar to the expansion that we described starting in section \ref{lco},
but there is one notable difference.  In section \ref{lco}, the quantum-corrected $G_2$-structure was described by a three-form $\phi'$ that
did not obey $d\phi'=0$; see eqns. (\ref{ai}) and (\ref{aaii}) for explicit formulas in lowest order.
By contrast, in the present analysis, the equation $d\h\phi=0$ is exact and only the equation $d\star\h\phi=0$ receives $\alpha'$ corrections.

Reexamining the formulas of section \ref{lco}, we see that at least in leading order, although $d\phi'\not=0$, one has $d(\phi'-\chi)=0$
 where $\chi$ is a locally-defined function of
$\phi'$ (this is shown in eqn.\ (\ref{eq:LeadingOrderChi}), where to the given order, $\phi$ can be replaced by $\phi'$).  Hence, although the   three-form $\phi'$ that determines the $\alpha'$-corrected metric is not closed,
it fails to be closed by a term that could be removed by a local change of variables, that is by the addition to $\phi'(y)$ at a point $y\in M$
of a function of $\phi'(y)$ and its derivatives
up to finite order.  It is logical to think that $\phi'-\chi$ in section \ref{lco} corresponds to $\h\phi$ in our present analysis.

We suggest that the relation between $\phi'$ and $\h\phi$ can be regarded as the relation between two different regularizations of the supersymmetric
$\sigma$-model with target $M$.  In section \ref{lco}, we used formulas that arise if one computes the effective action of ten-dimensional superstring
theory on a general ten-manifold $Z$ and then specializes to $Z=\R^3\times M$.  These formulas have ten-dimensional Poincar\'e covariance and (order
by order in $\alpha'$) locality,
but they do not have manifest spacetime supersymmetry.  The standard regularizations of the $\sigma$-model preserve Poincar\'e covariance
and locality  and lead to such formulas.  We will call such a regularization a 10-dimensional regularization.

By contrast, in section \ref{super}, we asked for a formalism that preserves three-dimensional covariance and locality and supersymmetry and seven-dimensional covariance and locality,
 but we did not assume ten-dimensional covariance.  If there is a regularization of the $\sigma$-model compatible with
these requirements, it is a different one than is customarily used in ten dimensions.  In such a regularization, one would expect $\h\phi$ (since
it is part of the superfield $\Cc$) to be the natural variable, rather than $\phi'$.
We will call this a $3\times 7$-dimensional supersymmetric regularization.

Two reasonable regularizations differ, order by order in perturbation theory, by a local change of variables.  Thus, the variable
$\h\phi$ used in a hypothetical $3\times 7$-dimensional supersymmetric regularization would be expected to differ by a local change of
variables from the variable $\phi'$ used in a standard
10-dimensional regularization.  This is the structure that we have found explicitly in the first non-trivial order,
and this encourages us to think that a $3\times 7$-dimensional supersymmetric regularization might exist, though we do not know how to construct one.

The significance might be as follows.  As we recalled in the introduction, in the case of compactification of Type II superstring theory on a Calabi-Yau
manifold, there is a nice $\sigma$-model argument \cite{NS} showing that supersymmetry is unbroken order by order in $\alpha'$.
We believe that a $3\times 7$-dimensional supersymmetric regularization, if it exists, might lead to a similar argument for $G_2$ manifolds.  All this
may have an analog for the problem we study in section \ref{spinseven}: a $2\times 8$-dimensional
supersymmetric regularization might provide a good framework for understanding $\alpha'$ corrections to compactification on an eight-manifold
of $\Spin(7)$ holonomy.

\subsubsection{Some Properties Of The Moduli Space}\label{someprop}

\def\M{{\mathcal M}}
Here we will describe  an interesting consequence of the relation (\ref{opolb}) between $\h\psi$ and $\h\phi$.

This consequence can be
deduced with no knowledge of the functional $K_0$ except its classical limit for $\alpha'\to 0$, which ensures that the  moduli
space  of superconformal $\sigma$-models with target $M$ goes over for $\alpha'\to 0$ so the corresponding classical moduli space of $G_2$
metrics.

Since $\h\phi$ is closed, its cohomology class $[\h\phi]$ defines an element of $H^3(M,\R)$.  Likewise, $\h\psi$
has a cohomology class $[\hat\psi]\in H^4(M,\R)$. In the classical limit, these go over to the cohomology classes of
 the familiar covariantly constant forms $\phi$ and $\psi=\star\phi$.
 Let $\M_0$ be the moduli space of $G_2$ metrics on $M$, modulo topologically trivial diffeomorphisms.
(This is a rough analog of the Teichmuller space of a Riemann surface.)   The classical moduli space $\M_0$ is conical (since a $G_2$ metric
can be rescaled by a positive constant).  The corresponding moduli space $\M$ of superconformally invariant
$\sigma$-models with target $M$ is not conical.  However, the fact that a classical $G_2$ metric
can be deformed order by order in $\alpha'$ to give a superconformally-invariant $\sigma$-model, and that this deformation is unique
up to a reparametrization of the variables in the $\sigma$-model, ensures that $\M$ looks like $\M_0$ near infinity.
This ensures that $\M$ inherits some properties of $\M_0$ (at least in the large volume region), so we will first state
some general properties of $\M_0$.

We define a map
$\varrho:\M\to Q=H^3(M,\R)\oplus H^4(M,\R)$
that takes a point in $\M$ to the point $[\h\phi]\oplus [\h\psi]\in  Q$.
This definition makes sense both in the classical theory at $\alpha'=0$ and (at least to all orders in $\alpha'$) in the quantum-corrected theory.
A basic result about classical $G_2$ manifolds is that if we set $\alpha'=0$, the map $\varrho$
 is locally an embedding.
Moreover, in the classical theory, $\varrho(\M)$
is middle-dimensional in $Q$ and can be parametrized locally by $[\h\phi]$ (with $[\h\psi]$ regarded as a function of $[\h\phi]$), as stated in Theorem 10.4.4 of \cite{joyce}.
All these statements, which are basic facts in the theory of classical $G_2$ manifolds,
 are stable under arbitrary small perturbations of the
map $\varrho:\M\to H^3(M,\R)\times H^4(M,\R)$.   So since these statements hold for the classical moduli space
$\M_0$, and given that $\M$ is asymptotic to $\M_0$ at infinity, they also hold for $\M$,
at least for sufficiently large volume.  (The definition we have given of the map
 $\varrho:
\M\to H^3(M,\R)\oplus H^4(M,\R)$
is only valid to all finite orders in $\alpha'$, but
we expect a natural such map to exist in general.  Perhaps it can be defined in superconformal field theory.  We expect the map
 $\varrho$
to be everywhere an immersion of a middle-dimensional submanifold, but the assertion
that $\varrho(\M)$ can be parametrized
locally by $[\h\phi]$ may be valid only near large volume.)

Now we come to a more delicate statement that does depend on the relation (\ref{opolb}).  For $\alpha'=0$, this statement is Proposition 10.4.5
in \cite{joyce}.   For a closed three-form $\h\phi$ and closed four-form $\h\psi$, the integral $\int_M\h\psi\wedge \h\phi$ depends only on the
cohomology classes $[\h\phi]$ and $[\h\psi]$.  We can use this integral to define a symplectic form $\varpi$
on   $Q=H^3(M,\R)\oplus H^4(M,\R)$:
\begin{equation}\label{torm} \varpi=\int_M\delta\h\phi\wedge \delta\h\psi. \end{equation}
The claim is that $\varrho(\M)$ is a Lagrangian submanifold of $Q$ (which means that $\varpi$ vanishes when restricted to $\varrho(\M)$).   We will explain this argument very explicitly.
We pick a basis of $H^3(M,\R)$ and a dual basis of $H^4(M,\R)$ and let $\h\phi^\lambda$ and $\h\psi_\lambda$ be the components
of $[\h\phi]$ and $[\h\psi]$ with respect to these bases. (For legibility of the following formulas, we prefer to write $\h\phi^\lambda$ and
$\h\psi_\lambda$ rather than $[\h\phi]^\lambda$ and $[\h\psi]_\lambda$.)    In these variables,
\begin{equation}\label{yorm}\varpi=\sum_\lambda d\h\phi^\lambda \wedge d\h\psi_\lambda. \end{equation}
(After reducing to the finite set of variables $\h\phi^\lambda$ and $\h\psi_\lambda$, we write the exterior derivative as $d$ rather than $\delta$.)
Once we restrict to $\M$, $\h\phi$ is uniquely determined by
is cohomology class $[\h\phi]$, so $K_0(\h\phi)$ can be regarded as a function of $[\h\phi]$ and hence of its components $\h\phi^\lambda$,  and
eqn.\ (\ref{opolb}) tells us that along $\M$,
\begin{equation}\label{worom} \h\psi_\lambda=\frac{\partial K_0}{\partial\h\phi^\lambda}. \end{equation}
It follows that
\begin{equation}\label{orom}d\h\psi_\lambda=\sum_\nu d\h\phi^\nu \frac{\partial^2K_0}{\partial \h\phi^\lambda\partial \h\phi^\nu }, \end{equation}
and hence when restricted to $\M$,
\begin{equation}\label{ozorm} \varpi= \sum_{\lambda,\nu}
d\h\phi^\lambda\wedge  d\h\phi^\nu \frac{\partial^2K_0}{\partial \h\phi^\lambda \partial \h\phi^\nu} =0, \end{equation}
where we use the fact that $\partial^2 K_0/\partial\h\phi^\lambda\partial\h\phi^\nu$ is symmetric in $\lambda$ and $\nu$.  This shows that
$\varrho(\M)$ is Lagrangian in $Q$.

We expect this conclusion to be valid exactly, not just to all finite orders in $\alpha'$.  The claim that
$K_0$ is a local functional of $\h\phi$ is valid only to all finite orders in $\alpha'$, but in the exact theory, we expect that there is a Kahler potential
on the space $\X$ that can be used as input to this analysis, leading to the same conclusion.

\section{$\Spin(7)$ Holonomy Manifolds}\label{spinseven}

In this section, we consider Type II superstring theory on a spacetime of the form $\RR^{1,1} \times M$,
where $M$ is a compact eight-dimensional manifold.   At the classical level, the condition
for  such a compactification to preserve supersymmetry on $\RR^{1,1}$, in the absence of fluxes,\footnote{For some results
on  the case with $G$-flux included
at the classical level (in the M-theory context and with $M$ restricted to have holonomy $SU(4)\subset \Spin(7)$),  see \cite{BeckerBecker}.  See \cite{Becker:2000jc} for an analysis of the analogous conditions for $\Spin(7)$.}
is that the holonomy group of $M$ should
be $\Spin(7)$ (or a subgroup thereof).   If the holonomy group is precisely $\Spin(7)$, which is the case we will concentrate on,
 then compactification on $M$ preserves $(1,1)$ supersymmetry on $\RR^{1,1}$  (Type IIA)
or $(0,2)$ supersymmetry (Type IIB).  Our goal is to investigate $\alpha'$ corrections to such compactifications.

An eight-manifold of $\Spin(7)$ holonomy admits a covariantly constant spinor field  $\eta$ of negative chirality,
which we normalize (up to sign) by setting
 $\eta^T \eta=1$.
$\Spin(7)$ and $G_2$ geometries share many common features when described in terms of
the covariantly constant spinor so we will be brief and collect technical details
in the appendix. We will have to explain, however, a few key differences between the two cases.

For any $\eta$ of negative chirality, the four-form
\begin{equation}\label{eqnoc}
\P_{abcd} = \et^T \Gamma _{abcd} \et,
\end{equation}
is anti-selfdual.
If  $\eta$ is covariantly constant,  then $\Phi$
 also covariantly constant and in particular
is closed and co-closed, $d\Phi=d\star\Phi=0$.

\def\mathbff{}
\subsection{Leading Order Correction}\label{lcoseven}

As in the $G_2$-holonomy case,
to describe corrections of order $\a'^3$ to the supersymmetry transformations,
we expand in a basis of real, antichiral spinors.  The negative chirality spinors
of $SO(8)$ decompose under $\Spin(7)$ as $1\oplus 7$, and the  two-forms
transform as $28=7\oplus 21$.  So we can take a basis of negative chirality spinors
 given by $\eta$ and $c^{ab}\Gamma _{ab}\eta$,
where $c^{ab}$ is an antisymmetric tensor transforming in the $\mathbff{7}$.
Hence the condition to have an unbroken supersymmetry with $\a'^3$ corrections included must
take the form of the existence of a spinor $\eta'=\eta+\O((\alpha')^3)$ satisfying
\begin{equation} \label{hi}
D_a'\eta'=A_a\eta+C_a^{\hph{a}bc}\Gamma _{bc}\eta,
\end{equation}
where $A_a$ and $C_a^{\hph{a}bc}$ are real tensors on $M$, proportional to $(\alpha')^3$, that are locally constructed from $\Phi$. Note that
$C_a^{\hph{a}bc}$ transforms in the $\mathbff{8}\otimes\mathbff{7}\cong\mathbff{8}\oplus\mathbff{48}$ of $\Spin(7)$.
Eqn. (\ref{hi}) is readily seen to imply that the one-form $A=A_a d y^a$ is $A=\hlf d\log (\eta'{}^T\eta')$, and hence if we rescale $\eta'$
so that $\eta'{}^T\eta'=1$  (this may not be the natural normalization for other purposes), it obeys eqn.\ (\ref{hi}) with $A_a=0$.
Actually, the standard formulas for the  $(\alpha')^3$ correction
do have $A_a=0$.

We define a corrected four-form
\begin{equation}\label{eqcor}
\P_{abcd}' = \et'^T \Gamma _{abcd}' \et'.
\end{equation}
The $\alpha'$-corrected condition (\ref{hi}) for unbroken supersymmetry is equivalent to the condition
\begin{equation}\label{ziz}
d \Phi' = \g
\end{equation}
for $\Phi'$, where we define
\begin{equation}
\g_{abcde}=-80C_{[a}^{\hph{[a}fg}g_{b|f|}\Phi_{cde]g}.
\end{equation}
Note that both the $\mathbff{8}$ and $\mathbff{48}$ pieces of $C$ appear in $\g$.  These are known as the torsion classes of the $\Spin(7)$-structure, and their data is equivalent to that contained in $C_a^{\hph{a}bc}$.

At leading order, from \cite{Lu:2004ng} we have
\begin{equation}
{C_a}^{bc}=-c \frac{\a'^3 }{4}\Phi_a^{\hphantom{a}qrs}\nabla_q Z_{rs}^{\hphantom{rs}bc}.
\end{equation}
Here $c$ is a constant, and $Z^{abcd}$ is given by
\begin{equation}
Z^{abcd}=\frac{1}{64 g }\e^{abe_1\cdots e_6}\e^{cdf_1\cdots f_6}R_{e_1e_2f_1f_2}R_{e_3e_4f_3f_4}R_{e_5e_6f_5f_6}.
\end{equation}
From here we find $\g=d\chi$ with
\begin{equation}
\chi_{abcd}=8c\Phi_{[abc}^{\hph{[abc}e}Z_{d]e}-12c\Phi_{[ab}^{\hph{[ab}ef}Z_{cd]ef}.
\end{equation}
So $\g $ is exact. As we describe next, this cohomology condition on $\g$ is necessary and
sufficient for the existence of a $\Spin(7)$-structure on $M$ that is close to the classical one
and obeys the $\alpha'$-corrected condition for supersymmetry, to this order.

Before trying to solve the eqn.\ (\ref{ziz}) for unbroken supersymmetry,
we must understand an important difference between a $G_2$-structure on a seven-manifold and a $\Spin(7)$-structure
on an eight-manifold.  At a given point $p$ in a seven-manifold, any three-form $\phi$ that obeys some inequalities
is invariant under a $G_2$ subgroup of the group $GL(7,\R)$ that acts on the tangent space at $p$.  (These inequalities
amount to saying that a metric can be defined by the formula of eqn.\ (\ref{metric1}).)  If $\phi$ obeys
the relevant inequalities,
we say it defines a $G_2$-structure at $p$.  A three-form $\phi$ on a seven-manifold $M$ that obeys the relevant conditions
everywhere on $M$ is said to define a $G_2$-structure on $M$; if $\phi$ has this property, then any three-form $\phi'$ that
is sufficiently close to $\phi$ does as well.  Matters are different in one dimension more.
At a given point $p$ in an eight-manifold $M$, a generic four-form $\Phi'$ is not invariant under a $\Spin(7)$ subgroup of the
group $GL(8,\R)$ that acts on the tangent space at $p$.  $\Phi'$ has $\Spin(7)$ symmetry if and only if
there is an orientation-preserving element of $GL(8,\R)$ that maps
$\Phi'$ to a standard fiducial four-form $\Phi_0$.
  Following ref. \cite{joyce}, section 10.5, we write  ${\cal A}_p M$ for the space of four-forms that obey this condition at a point $p\in M$,
  and ${\cal A}M$ for the space of four-forms that obey the condition for all $p\in M$.  From this description, clearly
$GL(8,\RR)$ acts transitively on ${\cal A}_p M$,
with the stabilizer group of a point being  $\Spin(7)$. So there is a one-to-one correspondence
between  ${\cal A}_p M$ and the coset $GL(8,\RR)/\Spin(7)$. Counting dimensions, we have
\begin{equation}
\mid GL(8,{\RR})/{\Spin(7)} \mid =43<70=\left( \begin{matrix}8 \\ 3 \end{matrix}\right)
=\left| \wedge  ^4 T^\star_p M \right|.
\end{equation}  So ${\cal A}_pM$ is not an open subset of the space $\wedge^4$ of all four-forms at $p$; it is of codimension $70-43=27.$
Under a $\Spin(7)$ subgroup  of $GL(8,\R)$, $\wedge^4$ decomposes as $1\oplus 7\oplus 35\oplus 27$.
If $\Phi$ defines a $\Spin(7)$-structure, then a deformation of $\Phi$ in the 1 represents a rescaling of $\Phi$,
preserving in each tangent space the $\Spin(7)$ subgroup of $GL(8,\R)$ that leaves $\Phi$ fixed.  A deformation in the $7\oplus  35$
represents a rotation of that $\Spin(7)$ subgroup inside $GL(8,\R)$.  But a deformation of $\Phi$ that is in the 27 cannot
be interpreted as a deformation of the $\Spin(7)$-structure.  Thus the tangent space to ${\cal A}_pM$ at the point corresponding
to $\Phi$ decomposes as $1\oplus 7\oplus 35$, and the condition that a deformation $\delta\Phi$ of $\Phi$ represents a deformation of
the $\Spin(7)$-structure of $M$ is that
\begin{equation}\label{thold}\pi_{27}(\delta\Phi)=0\end{equation}
 or equivalently
\begin{equation}\label{bhold}\delta\Phi\in \wedge ^4_1\oplus \wedge ^4_7\oplus \wedge ^4_{35}. \end{equation}

Explicitly, the meaning of the condition (\ref{bhold}) on $\delta\Phi$ is the following.
Comparing the definition of the unperturbed four-form $\Phi$ in eqn.\ (\ref{eqnoc}) to the definition
of the perturbed four-form $\Phi'$ in eqn.\ (\ref{eqcor}), we see that the perturbation $\Phi'$ has two sources:
{\it (i)} the perturbation from $\eta$ to $\eta'$; and {\it (ii)} the perturbation in the gamma matrices that appear in $\Gamma'_{abcd}$ in
eqn.\ (\ref{eqcor}).  Since we have set $\eta'{}^T\eta'=1$, the perturbation $\eta'-\eta$ transforms in the 7.  The perturbation in the
gamma matrices arise because of a perturbation in the metric $g$ of $M$.  The metric is a symmetric second rank tensor $g_{ab}$;
perturbations of $g_{ab}$ about a metric of $\Spin(7)$ holonomy transform as $1\oplus 35$.  Altogether, then, first order perturbations
of $\Phi$ should transform as $1\oplus 7\oplus 35$, with no contribution transforming as 27.

Now we expand
\begin{equation}\label{onold}\Phi'=\Phi+\delta\Phi,\end{equation}
and try to pick $\delta\Phi$ to satisfy $d(\delta\Phi)=\gamma$ and also $\pi_{27}(\delta\Phi)=0$.
If $\pi_{27}(\chi)=0$, we would simply take $\Phi'=\Phi+\chi$, and this would give the
deformed $\Spin(7)$-structure. However, $\pi_{27}(\chi)\neq 0$ and therefore a more elaborate discussion is required.
 We will try
\begin{equation}
\Phi'=\Phi+\lp 1-\pi_{27}\rp\lp\chi+dc\rp,
\end{equation}
for some three-form $c$.  By construction we have $\pi_{27}(\Phi')=0$, so it remains
only to show that we can find some globally-defined $c$ such that $d \P'=\g$, or
\begin{equation}\label{hx}
d\pi_{27}dc=-d\pi_{27}\chi.
\end{equation}
The decomposition of the space of three-forms  under $\Spin(7)$ is $\wedge^3=\wedge^3_8\oplus \wedge^3_{48}$.
A $c \in \wedge ^3_8$ corresponds to the change in $\Phi$ induced by an infinitesimal diffeomorphism. It does not contribute to eqn.\ \C{hx} since
$ 8 \otimes 8 \cong 1 \oplus 7 \oplus 21 \oplus 35$, so that $\pi_{27}d c=0$ for $c\in\wedge ^3_8$.
So we restrict to $c\in\wedge  ^3_{48}$. Similarly
the source $-d\pi_{27}\chi$ is also in $\wedge  ^5_{48}$ (dually to the decomposition of $\wedge^3$, one has $\wedge^5=\wedge^5_8\oplus \wedge^5_{48}$;
since $8$ does not appear in the decomposition of $8\otimes 27$, one has $d\pi_{27}\chi\in\wedge^5_{48}$).
Because of the decomposition $\wedge^4=\wedge^4_1\oplus \wedge^4_7\oplus \wedge^4_{27}\oplus \wedge^4_{35}$, we
have on four-forms $1=\pi_1+\pi_7+\pi_{27}+\pi_{35}$, so
\begin{equation}\label{tono}d\left(\pi_1+\pi_7+\pi_{27}+\pi_{35}\right)dc=0. \end{equation}
Hence the equation (\ref{hx}) that we are trying to solve is equivalent to
\begin{equation}\label{vono} d\left(\pi_1+\pi_7-\pi_{27}+\pi_{35}\right)d c = 2d\pi_{27}\chi. \end{equation}
For $c\in \wedge ^3_{48}$, $\pi_1(dc)=0$, since 1 does not appear in the decomposition of $8\otimes 48$.  So we can reverse
the sign of the $\pi_1$ term in eqn.\ (\ref{vono}).   We will show that we can solve eqn.\ (\ref{vono}) while also requiring
that $c$ is coclosed, $d^\dagger c=0$.  If this condition is satisfied, then $\pi_7(d c)=0$; indeed, 7 occurs only once in the
decomposition of $8\otimes 48$, so it occurs only once in the first derivatives of $c$, and this contribution is a multiple of $d^\dagger c$.
So if $d^\dagger c=0$, we can reverse the sign of the $\pi_7(d c)$ term in eqn.\ (\ref{vono}).  At this point, we use  the fact
that (since $\wedge^4_+=35 $ and $\wedge^4_-=1\oplus 7\oplus 27$) the Hodge $\star$ operator on four-forms is $\star=-\pi_1-\pi_7-\pi_{27}+\pi_{35}$.
Given this,
the equations that we have to satisfy can be written $d^\dagger d c=-2 d^\dagger \pi_{27}\chi$, $d^\dagger c=0$, or equivalently
\begin{equation}\label{zelf}\Delta c=-2 d^\dagger\pi_{27}\chi, ~~~  d^\dagger c=0,\end{equation}
where $\Delta=d^\dagger d+d d^\dagger$ is the Hodge-de Rham Laplacian.  By the general theory of the Hodge-de Rham Laplacian,
since  the source is orthogonal to the kernel of $\Delta$,
these  equations have a solution (unique up to the
possibility of adding a harmonic three-form to $c$), and the solution
can be taken to lie in $\wedge^3_{48}$ because the Hodge-de Rham Laplacian respects the decomposition $\wedge^3=\wedge^3_8\oplus
\wedge^3_{48}$ and the source lies in $\wedge^3_{48}$.

\subsection{All Orders In $\a'$}\label{allo}

To extend this result to all orders in $\a'$, we proceed as in section 2.2.
We expand
\begin{equation}\label{zold}\Phi'=\Phi+\sum_{k=3}^\infty (\alpha')^k \Phi_k.\end{equation}
As in the $G_2$ case, a solution will only exist, in a given order in the expansion, if a certain
closed form is actually exact.  In this case, in order $(\alpha')^k$, the condition that we will have
to satisfy is $d\Phi_k= \g[\Phi']|_k$, where $ \g[\Phi']|_k$ is a functional constructed locally from the lower
terms in the expansion for $\Phi'$.  Clearly a solution can only exist if $\g[\Phi']|_k$ is exact,
say $\g[\Phi']|_k=d\chi_k$ for some $\chi_k$.
In this case, we require
\begin{equation} \label{hxxi}
d\P_k =d\chi_k,\end{equation}
with also a constraint that ensures that $\Phi'\in {\mathcal A}M$.  If the constraint
were simply that $\pi_{27}\Phi_k=0$, then the problem of finding $\Phi_k$ would be isomorphic to the problem
already solved in leading order in  section \ref{lcoseven}, with a different source on the right hand side.
Actually, $\mathcal A M$ is a nonlinear space and the constraint  $\Phi'\in {\mathcal A}M$ is nonlinear in $\Phi'$.  As a result
the appropriate condition
on $\Phi_k$ is not that $\pi_{27}\Phi_k=0$, but that $\pi_{27}(\Phi_k)$ is a certain nonlinear function $\Theta_k$ of the $\Phi_n$, $n<k$.
After writing $\Phi_k=\Phi_k'+\Theta_k$ where $\pi_{27}(\Phi_k')=0$, eqn.\ (\ref{hxxi}) becomes
\begin{equation}\label{zeldix} d\Phi_k'=d(\chi_k-\Theta_k),\end{equation}
and now the problem is indeed isomorphic to the one that we have already studied.

\subsection{Interpretation In  Two- or Three-Dimensional  Field Theory}\label{intlow}

Here we will rather briefly interpret the results of sections \ref{lcoseven} and \ref{allo} in the language of supersymmetric field theory
in two or three dimensions.   As in our discussion of the $G_2$ case in section \ref{coco}, it will take more work to fully justify our proposal.

We can consider Type IIA superstring theory compactified to two dimensions on a $\Spin(7)$ manifold $M$, giving a two-dimensional
theory with $(1,1)$ supersymmetry, or Type IIB compactified on $M$, giving a two-dimensional theory with $(0,2)$ supersymmetry.
As  in section \ref{coco}, we will focus on the Type IIA case, and we also observe
that Type IIA compactification on $M$ to two-dimensions is similar to M-theory compactification on $M$ to three-dimensions,
with the $\alpha'$ expansion replaced by the $1/\br$ expansion.  In  M-theory, compactification on a $\Spin(7)$ manifold gives a three-dimensional
theory with $\N=1$ supersymmetry (two supercharges).   Our discussion applies to each of these cases.

Two-dimensional theories with $(1,1)$ supersymmetry and three-dimensional theories with $\N=1$ supersymmetric can be conveniently
formulated in terms of a superspace with two or three bosonic coordinates $x^\mu$ and two fermionic coordinates $\theta^\alpha$, $\alpha=1,2$.
The superfields that will be important in our analysis are unconstrained functions  $\Lambda(x^\mu,\theta^\alpha)$ on superspace
(these are often called scalar superfields, but we will be more precise in our terminology below).  For the purposes of finding supersymmetric vacuum states, the most important supersymmetric interaction is the
superpotential.  This is a function $W(\Lambda)$ such that the condition for a supersymmetric vacuum is $dW=0$.

In either Type IIA or M-theory compactification on a $\Spin(7)$ manifold, there is a discrete symmetry $\tau$ under which the superpotential is odd.
In Type IIA, one can take $\tau$  to be the operation $(-1)^{F_L}$ that reverses the sign of left-moving worldsheet fermions.
In M-theory, one can take $\tau$ to be a reflection of one of the uncompactified directions accompanied by a sign change of the three-form field $C$.
There are therefore two kinds of superfield: we call $\Lambda$ a scalar superfield if
its bottom component is even under $\tau$, and a pseudoscalar superfield if its bottom component is odd under $\tau$.

We will only analyze supersymmetric vacua that are $\tau$-invariant (this means that we omit vacua with fluxes of the field $G=d C$,
as have been studied in \cite{BeckerBecker}).  Let $S_I$ be the scalar superfields and $T_J$ the pseudoscalar superfields.
Since the superpotential is odd under $\tau$, it can be expanded in powers of the $T$'s with only odd order terms appearing:
\begin{equation}\label{ohno}W=\sum_IT_I F_I(S_J)+\O(T^3). \end{equation}  Here the $F_I$ are in general completely arbitrary functions of the
$S_J$.
In a globally supersymmetric theory, the $\tau$-invariant supersymmetric states correspond to solutions of $dW=0$ with also $T=0$.  Clearly the necessary
condition is simply that
\begin{equation}\label{bono} F_I(S_J)=0 \end{equation}
for all $I$.  After coupling to supergravity, one also wants $W=0$ to get a supersymmetric vacuum in Minkowski spacetime; clearly in a theory of this kind,
this is an immediate consequence of setting $T=0$.

In M-theory on $\R^3\times M$ (or similarly in Type IIA on $\R^2\times M$), the obvious pseudoscalar fields are obtained by taking all indices
of the three-form field $C$ to be tangent to $M$.  This gives us a field $C_{abc}(x,y)$, which as in section \ref{coco} we regard as a pseudoscalar
field on $\R^3$ that is also a three-form on $M$.  We expect  $C_{abc}(x,y)$ to be the bottom component of a superfield that we will, for brevity,
also call $C_{abc}$.  We will assume that to describe the theory in a way that has manifest covariance, locality and supersymmetry
along $\R^3$ or $\R^2$ and manifest covariance and locality along $M$, we should also introduce a scalar superfield\footnote{How is $\hat\Phi$
constructed in terms of the usual degrees of freedom of supergravity?  As explained
in section \ref{lcoseven}, a perturbation in $\hat\Phi$ can be decomposed under $\Spin(7)$ as $1\oplus 7\oplus 35$, while a perturbation in the
metric $g$ of $M$ can be decomposed as $1\oplus 35$.  What is the proper interpretation of the 7 contribution in $\hat\Phi$?
In section \ref{super}, it was suggested that the answer to the analogous question for $G_2$ (in the M-theory case) arises from
scalar fields obtained by dualizing $C_{\mu \nu a}$.  For $\Spin(7)$ an analog is to consider scalar fields obtained by dualizing $C_{\mu ab}$,
which transforms as $7\oplus 21$.  The 7 would hypothetically complete the construction of $\hat\Phi$ and the 21 would represent additional
fields that would be included in a more complete description.}
 $\hat\Phi$ that is a four-form
on $M$ constrained to take values in ${\mathcal A}M$.

Because of the usual gauge-invariance $C\to C+d\Lambda$
for a two-form $\Lambda$ on $M$, the superpotential $W$ depends on $C$ only through its field strength $G=d C$, where here we consider only the
part of $G$ that is a four-form on $M$ (and a pseudoscalar function on $\R^3$).  We expand the superpotential in powers of $G$; as in the last
paragraph only odd powers appear and only the linear term is important for understanding supersymmetric vacua at $G=0$.  Thus the general
form of the  superpotential is
\begin{equation}\label{mondee} W=\int_M G\wedge P(\hat\Phi) +\O(G^3) \end{equation}
where to any finite order in $\alpha'$, $P$ is a local functional of $\hat\Phi$.
The condition for a critical point at $G=0$ is simply
\begin{equation}\label{wondee}d P(\hat\Phi)=0. \end{equation}

In the classical limit $\alpha'\to 0$, we must have simply $P(\hat\Phi)=\hat\Phi$, so that eqn.\ (\ref{wondee}) reduces to the classical condition
$d\hat\Phi=0$ which (with the constraint that $\hat\Phi$ is valued in $\mathcal AM$) is equivalent to $\Spin(7)$ holonomy.
But now we can easily go to higher orders.  When we expand
\begin{equation}\label{zoldo}\hat\Phi=\Phi+\sum_{k=3}^\infty (\alpha')^k \Phi_k,\end{equation}
the equation (\ref{wondee}) will give, in each order, an equation of the general
form
\begin{equation}\label{poldo}d\Phi_k=d\chi_k \end{equation}
where $\chi_k$ is a nonlinear function of the $\Phi_n$ for $n<k$.  There will also be a constraint that determines $\pi_{27}(\Phi_k)$ in terms
of the previous $\Phi_n$'s.  As we have seen, it is always possible to solve conditions of this form.  The main point is that the supersymmetric
structure implies that the five-forms that were called $\left.\gamma[\Phi']\right|_k$ in section
\ref{allo} are always exact.  In fact, they are always $d\chi_k$, where $\chi_k$ is a local function of the $\Phi_n$'s with $n<k$.

\appendix
\section{Background On Manifolds Of Exceptional Holonomy}

In this appendix we review some facts
which have been used in the main body of the paper. These facts are collected for the
convenience of the reader.

Before getting into details on $G_2$ or $\Spin(7)$, recall that the  standard inner product of forms is defined by
\begin{equation}
\left\langle\chi,\xi\right\rangle=\frac{1}{p!}\int d^d x \sqrt{g}\chi^{a_1\cdots a_p}\xi_{a_1\cdots a_p}.
\end{equation}
This inner product respects the decomposition of forms into $G_2$ and $\Spin(7)$ representations,
so for distinct irreducible representations $\mathbf{r}$ and $\mathbf{s}$
\begin{equation}
\left\langle\pi_r(\chi),\pi_s(\xi)\right\rangle=0.
\end{equation}
We use $d^\dagger=(-1)^{p(d+1-p)}\star d\star$ for the adjoint of the exterior derivative acting on
 $p$-forms. With this definition,
\begin{equation}
\label{eq:DerivativeAdjoint}
\left\langle\chi,d\xi\right\rangle=\left\langle d^\dagger \chi,\xi\right\rangle.
\end{equation}

\subsection{$G_2$}

\subsubsection{Spinor Conventions}

We use a basis in which the gamma matrices $\Gamma _a$ are purely imaginary antisymmetric matrices satisfying $\{\Gamma _a,\Gamma _b\}=2g_{ab}$.  The Clifford algebra is spanned by real symmetric matrices $\{ {\bf 1}, i \Gamma _{abc} \}$
and real anti-symmetric matrices $\{ i \Gamma _a, \Gamma _{ab} \}$. The eight spinors $\{\eta, i\Gamma _a\eta\}$,
are a basis.  The completeness of this basis says
\begin{equation}\label{comp}
\Gamma ^a\eta\eta^T\Gamma _a+\eta\eta^T={\bf 1}.
\end{equation}
Moreover, in $7d$ we impose
\begin{equation}
{1\o 7!} \sum_{a_1,\dots,a_7}\ve_{a_1\dots a_7} \Gamma ^{a_1 \dots a_7 } = -i,
\end{equation}
where the Levi-Civita symbol $\ve_{a_1\dots a_7}$ is a tensor.

Define
\begin{equation}
\vp_{abc} = i \et^T \Gamma _{abc} \et,
\qquad
\ps_{abcd} = {1\o 3!} \ve_{abcd klm} \vp ^{klm} = \et^T \Gamma _{abcd} \et.
\end{equation}
Any real spinor can be expanded in the above basis. In particular
\begin{equation}
\begin{split}
\Gamma _{ab}\eta & =-i\vp_{ab}^{\hph{ab}c}\Gamma _c\eta,\\
i\Gamma _{abc}\eta & =\vp_{abc}\eta-i\psi_{abc}^{\hph{abc}k}\Gamma _k\eta.
\end{split}
\end{equation}
Using \C{comp} one derives some useful identities
\begin{equation}
\begin{split}
{\vp_{ab}}^k \vp_{cdk } & = 2 g_{a [c} g_{d] b} - \ps_{abcd},  \\
\ps_{abc k} \vp^{dek} & = 6 \d^{[d}_{[a} {\vp_{bc]}}^{e]}, \\
{\ps_{abc}}^ k{\ps^{def}}_k & = - 9 \d^{[d}_{[a} {\ps_{bc]}}^{ef]} + 6 \d_{[abc]}^{def} - \vp_{abc} \vp^{def},
\end{split}
\end{equation}
together with additional identities obtained by contraction.

\subsubsection{Metric}

Given a $G_2$-structure $\vp$ the above relations can be used to obtain
\begin{equation}\label{metric1}
g_{ab}=-\frac{1}{144}\e^{ijklmnp}\vp_{aij}\vp_{bkl}\vp_{mnp}.
\end{equation}
Given a $G_2$-structure $\vp$ this can be used as the defining equation for the metric $g_{ab}$. Indeed, the epsilon tensor
takes values\footnote{\label{welf}Given a choice of $\vp$, the sign of the epsilon tensor or equivalently the orientation of $M$
is determined to make the metric
defined in (\ref{metric1}) positive.  However, without changing the metric, we could reverse the sign of $\vp$ and also the orientation
of $M$.  Thus a $G_2$ manifold does not have a preferred orientation.}
$\pm 1/\sqrt{g}$ and so taking the determinant of eqn.\C{metric1}
we can solve for $g$ in terms of $\vp$ and this in turn lets us
write the metric $g_{ab}$ in terms of $\vp$ only.
We can then consider $\vp$
to be the fundamental object from which the metric, Riemann tensor, covariant derivatives and $\ps = \star \vp$ are obtained.

\subsubsection{Decomposition Of Differential Forms Into Irreducible Representations Of $G_2$}

Using the fact that the tangent and cotangent spaces at points of $M$ transform as the fundamental seven-dimensional representation of $G_2$, one can derive the transformations of $p$-forms (living in $\wedge ^p\cong\wedge ^p(T^\ast M)$),
\begin{equation}
\begin{split}
\wedge ^1 &\cong \wedge ^1_7,\\
\wedge ^2 &\cong \wedge ^2_7\oplus\wedge ^2_{14},\\
\wedge ^3 &\cong \wedge ^3_1\oplus\wedge ^3_7\oplus\wedge ^3_{27}.
\end{split}
\end{equation}
We also have $\wedge ^{7-p}\cong\wedge ^p$. We list the projections

\begin{itemize}

\item[$\tria$] for $\wedge ^2 \cong \wedge ^2_7 \oplus \wedge ^2_{14}$

\begin{equation}
\begin{split}
\lp\pi_7\a\rp_{ab} &= \frac{1}{3}\a_{ab}-\frac{1}{6}\psi_{ab}^{\hphantom{ab}cd}\a_{cd},\\
\lp\pi_{14}\a\rp_{ab} &= \frac{2}{3}\a_{ab}+\frac{1}{6}\psi_{ab}^{\hphantom{ab}cd}\a_{cd}.
\end{split}
\end{equation}

\item[$\tria$] for $\wedge ^3 \cong \wedge ^3_1\oplus \wedge ^3_7 \oplus \wedge ^3_{27}$
\begin{equation}
\begin{split}
\lp\pi_1\beta\rp_{abc} =&  \frac{1}{42}\vp_{abc} \vp^{def}\beta_{def},\\
\lp\pi_7\beta\rp_{abc} =& \frac{1}{4}\beta_{abc}-\frac{3}{8}\psi_{[ab}^{\hphantom{[ab}de}\beta_{c]de}-\frac{1}{24}\vp_{abc}\vp^{def}\beta_{def},\\
\lp\pi_{27}\beta\rp_{abc} =& \frac{3}{4}\beta_{abc}+\frac{3}{8}\psi_{[ab}^{\hphantom{[ab}de}\beta_{c]de}+\frac{1}{56}\vp_{abc}\vp^{def}\beta_{def}.
\end{split}
\end{equation}

\end{itemize}

\subsubsection{Deformations Of $G_2$-Structures}

The deformed structure $\vp'=\vp+\d\vp$ will give rise to deformations in the metric, $g'=g+\d g$, and the four-form, $\psi'=\psi+\d\psi$.  Plugging these into the contraction
\begin{equation}
{\vp_a}^{cd} \vp_{bcd} = 6 g_{ab},
\end{equation}
one can derive that to first order in the deformation $\d\vp$ we have
\begin{equation}
\d g_{ab}=-\frac{1}{18}g_{ab}\vp^{cde}\d\vp_{cde}+\hlf\vp_{(a}^{\hph{(a}cd}\d\vp_{b)cd},
\end{equation}
and using this metric to construct the Hodge star we get
\begin{equation}
\d\psi_{abcd}=-\frac{1}{9}\psi_{abcd}\vp^{efg}\d\vp_{efg}-\frac{1}{3}\vp_{[abc}\psi_{d]}^{\hph{d]}efg}\d\vp_{efg}-6\vp_{[ab}^{\hph{[ab}e}\d\vp_{cd]e}.
\end{equation}
In both of these expressions, indices are raised with the undeformed metric $g^{ab}$.

In terms of the pieces of $\d\vp$ which transform in different representations of $G_2$ (the copy of $G_2$ which leaves the original $\vp$ invariant), we can rewrite these first order deformations as
\begin{equation}
\d g_{ab}=\vp_{(a}^{\hph{(a}cd}\lp\frac{1}{9}\pi_1(\d\vp)+\hlf\pi_{27}(\d\vp)\rp_{b)cd},
\end{equation}
\begin{equation}
\d\psi=\frac{4}{3}\star \pi_1(\d\vp)+\star \pi_7(\d\vp)-\star \pi_{27}(\d\vp).
\end{equation}

\subsubsection{The Torsion Forms}
\label{app:G2Torsion}

Given a $G_2$-structure $\vp'$, its exterior derivative $d \vp'$ is a four-form and
the exterior derivative of the dual $d\psi'= d\star' \vp'$ is a five-form. These forms
can be decomposed into irreducible representations of $G_2$. Following proposition 1 of ref. \cite{Bryant:2005mz}, we
write
\begin{equation}
\begin{split}
d \vp' & = \t_0 \ps' + 3 \t_1 \wedge \vp' + \star' \t_3, \\
d \ps' & = 4 \tilde \t_1 \wedge \ps' + \star' \t_2,
\end{split}
\end{equation}
where $\t_3 \in \wedge^3_{27}$ and $\t_2 \in \wedge^2_{14}$. It was mentioned in ref. \cite{Bryant:2005mz}
that the projections of $d \vp'$ and $d \psi'$ onto $\wedge^4_7$ and $\wedge^5_7$ are closely related and in
particular
$\tilde \t_1 = \t_1$. The basic reason is that for any given $G_2$-structure
the following identity holds (eqn.\ (3.8) of ref. \cite{Bryant:2005mz})
\begin{equation} \label{ki}
(d\ps')_{abcde} {\psi'}^{bcde} = 4  (d {\vp'})_{abcd} {\vp'}^{bcd};
\end{equation}
this can be shown using the definition $\psi'= \star' \vp' $.
Here we note a consequence of this result which we use in the main body of the paper.
If we consider a perturbation about a $G_2$ holonomy space with $G_2$-structure $\vp$ by setting
$\vp' = \vp + \d \vp$, then to first order in fluctuations eqn.\ \C{ki} becomes
 \begin{equation} \label{kii}
(d  \ps')_{abcde} \psi^{bcde} = 4  (d {\vp'})_{abcd} {\vp}^{bcd}=
\left[d\star \left({4\o 3} \pi_1 + \pi_7 -\pi_{27}  \right)\d \vp \right]_{abcde} {\psi}^{bcde}.
\end{equation}
In section \ref{existence}, to first order in $\a'$, we solve the conditions
$ d \vp' = d \chi$ and $d \ps' = d \xi$ for globally-defined $\chi$ and $\xi$. In this case eqn.\ \C{kii} implies
 \begin{equation} \label{kiii}
(d \xi )_{abcde} \psi^{bcde} = \left[d\star \left({4\o 3} \pi_1 + \pi_7 -\pi_{27}  \right)\chi \right]_{abcde} {\psi}^{bcde} .
\end{equation}
This is equivalent to the vanishing of the $\wedge^2_7$ projection of $d^\dagger \r$, which is defined in eqn.\ \C{zi}.

A similar argument can be used to show that $d^\dagger \r_n\in \wedge_{14}^2$ in section 2.3. In this case we take
\begin{equation}
\vp' = \vp + \sum_{i=3}^{n-1} \vp_i + \vp_n = \tilde \vp + \delta \vp.
\end{equation}
So $\tilde \vp$ is the $G_2$-structure up to order $n-1$ in $\a'$. We consider a small perturbation of order $n$ around this $G_2$-structure
$\d \vp= \vp_n$ .
Then
\begin{equation}
\psi' =\tilde \psi+ \d \psi,
\end{equation}
where $ \tilde \psi = \tilde \star \tilde \vp$. Note that $\tilde \vp$ is at most of order $n$ while $\psi'$ could, in principle, receive
contributions to all orders in $\a'$ since it it a non-linear functional of $\tilde \vp$. So
\begin{equation} \label{exactaiii}
\psi'\mid_n = \tilde \star \tilde \vp\mid_n + \d \psi\mid_n.
\end{equation}
If we then expand eqn.\ \C{ki} about $\tilde \vp$ and use the fact that \C{ki} is valid with primed quantities substituted by tilde quantities
we derive
\begin{equation} \label{exactiv}
\left[ d (\d \psi)\right]_{abcde} \psi^{bcde}\mid _n = 4 \left[ d (\d \vp) \right]_{abcd} \vp^{bcd} \mid_n,
\end{equation}
at order $n$ in $\a'$. According to \C{exactaiii}
\begin{equation}
d( \d \psi)\mid_n= d \psi'\mid_n  - d(\tilde \star \tilde \vp)\mid_n .
\end{equation}
Now, locally we can always solve \C{eq:GeneralG2GravitinoVar} to find $\eta'$ such that the associated forms $\vp'$ and $\psi'$ satisfy \C{exactai}, but a priori we may not be able to extend $\eta'$ to a global solution.  Note that the local solution for $\vp'$ can always be written in the form $\chi_n+db_n$, where now $b_n$ may not be globally defined.  By substituting this local solution for $\psi'$ above, we obtain
\begin{equation}
\left( d\xi_n - d (\tilde \star \tilde \vp)\mid_n\right)_{abcde} \psi^{bcde} = 4 \left( d \chi_n \right)_{abcd} \vp^{bcd} ,
\end{equation}
which is equivalent to the vanishes of the $7$ part of the source $d^\dagger \r_n$.  Finally, since $b_n$ can be shown to drop out of this relation completely, the result will hold true for any valid solution.

\subsubsection{Some Useful Identities}

If $M$ is a $G_2$ holonomy manifold, a few useful properties can be derived. Defining
\begin{equation}
\lp L\cdot\lambda\rp_{abc}=\psi_{abc}^{\hph{abc}d}\lambda_d, \qquad \l\in \wedge ^1,
\end{equation}
we have
\begin{equation}
\label{eq:OneFormIdentity}
d^\dagger L\cdot\lambda=\lp 2\pi_7-\pi_{14}\rp d\lambda.
\end{equation}
For any two-forms
\begin{equation}
\label{eq:TwoFormRelation}
\pi_7(db)=-\frac{1}{4}L\cdot d^\dagger b, \qquad b \in \wedge ^2_{14},
\end{equation}
and
\begin{equation}
\label{eq:7DropsOut}
d^\dagger \lp\frac{4}{3}\pi_1+\pi_7-\pi_{27}\rp d \a=0, \qquad \a \in \wedge ^2_7.
\end{equation}
Another useful identity is
\begin{equation} \label{im1}
\pi_1 (db)=0, \qquad \forall b \in \wedge ^2_{14}.
\end{equation}

\subsection{$\Spin(7)$}

\subsubsection{Spinor Conventions}

For an eight-manifold with $\Spin(7)$-structure,
we have a nowhere-vanishing real spinor $\eta$.  We will choose conventions in which $\eta$ is antichiral,
\begin{equation}
\Gamma _9\eta=-\eta,
\end{equation}
where
\begin{equation}
\Gamma _9=\frac{1}{\sqrt{g}}\Gamma _1\Gamma _2\cdots\Gamma _8,
\end{equation}
and $\Gamma _i$ are pure imaginary antisymmetric $16\times 16$ gamma matrices for $SO(8)$.  We can normalize $\eta$ so that
\begin{equation}
\eta^T\eta=1,
\end{equation}
and we also have properties
\begin{equation}
\Gamma _a\eta\eta^T\Gamma ^a=\Pi_+,\qquad\eta\eta^T-\frac{1}{8}\Gamma _{ab}\eta\eta^T\Gamma ^{ab}=\Pi_-,
\end{equation}
where
\begin{equation}
\Pi_\pm=\hlf\lp 1_{16\times 16}\pm\Gamma _9\rp.
\end{equation}

Define
\begin{equation}
\Phi_{abcd}=\eta^T\Gamma _{abcd}\eta.
\end{equation}
Note that since
\begin{equation}
\frac{\sqrt{g}}{4!}\e_{abcdefgh}\Gamma ^{efgh}=\Gamma _9\Gamma _{abcd},
\end{equation}
we have $\star \Phi=-\Phi$.
We can also derive
\begin{equation}
\label{eq:PhiContraction1}
\Phi^{abcg}\Phi_{defg}=6\d_{[d}^{[a}\d_e^b\d_{f]}^{c]}-9\d_{[d}^{[a}\Phi^{bc]}_{\hphantom{bc]}ef]},
\end{equation}
and its contractions.

\subsubsection{Decomposition Of Differential Forms Into Irreducible Representations Of $\Spin(7)$}

Under $\Spin(7)$, the spaces of differential forms decompose as
\begin{equation}
\begin{split}
\wedge  ^0 &\cong \wedge  ^0_1,\\
\wedge  ^1 &\cong \wedge  ^1_8,\\
\wedge  ^2 &\cong \wedge  ^2_7\oplus\wedge  ^2_{21},\\
\wedge  ^3 &\cong \wedge  ^3_8\oplus\wedge  ^3_{48},\\
\wedge  ^4 &\cong \wedge  ^4_1\oplus\wedge  ^4_7\oplus\wedge  ^4_{27}\oplus\wedge  ^4_{35},
\end{split}
\end{equation}
and $\wedge  ^{8-n}\cong\wedge  ^n$.  We list the projections

\begin{itemize}

\item[$\tria$] for $\wedge  ^2\cong\wedge  ^2_7\oplus\wedge  ^2_{21}$, we have
\begin{equation}
\begin{split}
\lp\pi_7\a \rp_{ab} &= \frac{1}{4}\a _{ab}-\frac{1}{8}\Phi_{ab}^{\hphantom{ab}cd}\a _{cd},\\
\lp\pi_{21}\a \rp_{ab} &= \frac{3}{4}\a _{ab}+\frac{1}{8}\Phi_{ab}^{\hphantom{ab}cd}\a _{cd}.
\end{split}
\end{equation}

\item[$\tria$] for $\wedge  ^3\cong\wedge  ^3_8\oplus\wedge  ^3_{48}$,
\begin{equation}
\begin{split}
\lp\pi_8\beta\rp_{abc} =& \frac{1}{7}\beta_{abc}-\frac{3}{14}\Phi_{[ab}^{\hphantom{[ab}de}\beta_{c]de},\\
\lp\pi_{48}\beta\rp_{abc} =& \frac{6}{7}\beta_{abc}+\frac{3}{14}\Phi_{[ab}^{\hphantom{[ab}de}\beta_{c]de}.
\end{split}
\end{equation}

\item[$\tria$] and for $\wedge  ^4\cong\wedge  ^4_1\oplus\wedge  ^4_7\oplus\wedge  ^4_{27}\oplus\wedge  ^4_{35}$,
\begin{equation}
\begin{split}
\lp\pi_1\g\rp_{abcd} &= \frac{1}{336}\Phi_{abcd}\Phi^{efgh}\g_{efgh},\\
\lp\pi_7\g\rp_{abcd} &= \frac{1}{8}\g_{abcd}-\frac{3}{16}\Phi_{[ab}^{\hphantom{[ab}ef}\g_{cd]ef}-\frac{1}{48}\Phi_{[abc}^{\hphantom{[abc}e}\Phi_{d]}^{\hphantom{d]}fgh}\g_{efgh},\\
\lp\pi_{27}\g\rp_{abcd} &= \frac{3}{8}\g_{abcd}+\frac{15}{16}\Phi_{[ab}^{\hphantom{[ab}ef}\g_{cd]ef}+\frac{1}{56}\Phi_{abcd}\Phi^{efgh}\g_{efgh}-\frac{1}{16}\Phi_{[abc}^{\hphantom{[abc}e}\Phi_{d]}^{\hphantom{d]}fgh}\g_{efgh},\\
\lp\pi_{35}\g\rp_{abcd} &= \hlf\g_{abcd}-\frac{3}{4}\Phi_{[ab}^{\hphantom{[ab}ef}\g_{cd]ef}-\frac{1}{48}\Phi_{abcd}\Phi^{efgh}\g_{efgh}+\frac{1}{12}\Phi_{[abc}^{\hphantom{[abc}e}\Phi_{d]}^{\hphantom{d]}fgh}\g_{efgh}.
\end{split}
\end{equation}

\end{itemize}

The space of four-forms decomposes into self-dual forms $\wedge  ^4_+$ and anti-self-dual forms $\wedge  ^4_-$
\begin{equation}
\label{eq:SelfDualDecomp}
\wedge  ^4_+\cong\wedge  ^4_{35},\qquad\wedge  ^4_-\cong\wedge  ^4_1\oplus\wedge  ^4_7\oplus\wedge  ^4_{27}.
\end{equation}

\subsubsection{Deformations Of $\Spin(7)$-Structures}

The metric of a $\Spin(7)$-structure has been derived in ref. \cite{kari1}. Here
we require the result for the deformations of the metric.
The deformed structure $\Phi'=\Phi+\d\Phi$ will give rise to a metric deformation
\begin{equation}
\begin{split}
\d g_{ab} & =-\frac{1}{112}g_{ab}\Phi^{cdef}\d\Phi_{cdef}+\frac{1}{12}\Phi_{(a}^{\hph{(a}cde}\d\Phi_{b)cde}\\
& =\Phi_{(a}^{\hph{(a}cde}\lp\frac{1}{21}\pi_1(\d\Phi)+\frac{1}{12}\pi_{35}(\d\Phi)\rp_{b)cde}.
\end{split}
\end{equation}
With this metric, (\ref{eq:PhiContraction1}) continues to hold (at leading order) with $\Phi$ replaced by $\Phi'$ and $g$ by $g'$.

\subsubsection{The $L$ Operator}

On a $\Spin(7)$ manifold we define a map $L:\wedge^p \to \wedge^{p+2}$ by
\begin{equation}
(L \om)_{a_1 \dots a_{p+2}} = {\P_{[a_1a_2 a_3}}^p \om_{a_4 \dots a_{p+2}] p}.
\end{equation}
On forms in irreducible representations of $G_2$, $L$ does not change the representation. Moreover, $L$ is invertible on its image.
This leads to the isomorphisms
\begin{equation}
\wedge^1_8 \cong \wedge_8^3 \cong \wedge_8^5 \cong \wedge_8^7, \qquad \wedge_7^2 \cong \wedge_7^4 \cong \wedge_7^6, \qquad \wedge^3_{48} \cong \wedge^5_{48}.
\end{equation}
The kernel is
\begin{equation}
L (\wedge^2_{21}) = 0 , \qquad L(\wedge_1^4 \oplus \wedge^4_{27} \oplus \wedge^4_{35}) = 0 , \qquad L(\wedge^5_{48})=0,\qquad  L(\wedge^p)=0, \quad p \geq 6.
\end{equation}

There are potentially three ways of constructing a $(p+1)$-form by differentiating a $p$-form $\om$;
we can make $d\om$, $d^\dagger L\om$, or $Ld^\dagger\om$.  If $\om$ transforms in an irreducible representation $r$,
then each of these must transform in representations contained in the product
$8\otimes r$. If some given irreducible representation $s$ only
occurs once in the decomposition of $8 \otimes  r $, then it means that the forms $\pi_s (d\om)$,
$\pi_s (d^\dagger  L\om)$, and $\pi_s (L d^\dagger\om)$ must all be proportional to each other.
We use this to derive some useful identities:

\begin{itemize}

\item[$\tria$] if $\om \in \wedge^1_8$, we 
find the identities
\begin{equation}
\pi_7(d\om)=\frac{1}{3}\pi_7d^\dagger (L\om ),\qquad\pi_{21}(d\om)=-\pi_{21}d^\dagger (L\om).
\end{equation}

\item[$\tria$] for $\om \in \wedge^2_7$

\begin{equation}
Ld^\dagger\om=\frac{7}{3}\pi_8(d\om)=-\frac{7}{4}\pi_8 (d^\dagger L\om),
\end{equation}
and
\begin{equation}
\pi_{48}(d\om)=\pi_{48}(d^\dagger L\om),\quad\pi_{48}(Ld^\dagger\om)=0.
\end{equation}

\item[$\tria$] for $\om\in\wedge^2_{21}$, we have $d^\dagger L\om=0$ (since $L\om=0$) and
\begin{equation}
Ld^\dagger\om=-7\pi_8(d\om),
\end{equation}
and
\begin{equation}
\pi_{48}(Ld^\dagger\om)=0.
\end{equation}

\item[$\tria$]
similar relations can be derived for higher degree forms.  The only other facts we will need are that $\pi_{27}(d\om)=0$ for $\om\in\wedge^3_8$ and for $\om\in\wedge^3_{48}$, we have $\pi_1 (d\om)=0$ and
\begin{equation}
Ld^\dagger\om=-4\pi_7(d\om).
\end{equation}

\end{itemize}

Starting with the standard Hodge decomposition,
the above identities can be used to derive the following decompositions

\begin{itemize}

\item[$\tria$] for any $\a \in \wedge ^2_{21}$, there exists a one-form $\s$ and a co-closed
three-form $\r\in \wedge ^2_{21}$ such that
\begin{equation}
\a = \pi_{21} ( d \s) + \r.
\end{equation}

\item[$\tria$] for any $\xi \in \wedge ^3_{48}$, there exists a two-form $\mu$ and
co-closed three-form $\n \in \wedge ^3_{48}$ such that
\begin{equation} \label{zxi}
\xi = \pi_{48} (d \mu) + \nu.
\end{equation}

\end{itemize}

\section*{Acknowledgement}\addcontentsline{toc}{section}{Acknowledgement} The research of K.B. was supported by a grant from the Ambrose Monell Foundation, NSF grants PHY-0906222, PHY-1214344 and NSF Focused Research Grant DMS-1159404.
D.~R. was supported by funding from the European Research Council, ERC grant agreement no. 268088-EMERGRAV, NSF grant PHY-0906222 and by the
Mitchell Institute for Fundamental Physics and Astronomy. Research of E.W. was partly supported by NSF Grant PHY-1314311.
K.B. thanks the Institute for Advanced Study and the Banff International Research Station for hospitality during
different stages of this work and would like to thank Melanie Becker for discussions.  D.~R. would like to thank the Institute for Advanced Study, the Banff International Research Station, and the Mitchell Institute for hospitality, and would like to thank Jan de Boer, Ruben Minasian, and Savdeep Sethi, and Erik Verlinde for discussions.

\newpage

\providecommand{\href}[2]{#2}\begingroup\raggedright
\endgroup

\begin{thebibliography}{10}

\bibitem{gvz}
M. T. Grisaru, A. E. M. van de Ven, and D. Zanon, ``Four Loop Beta Function For The $\N=1$ and $\N=2$ Nonlinear Sigma Model
In Two Dimensions,''  Phys.Lett. {\bf B173} (1986) 423.

\bibitem{sv}
S. Shatashvili and C. Vafa, ``Superstrings And Manifolds Of Exceptional Holonomy,'' Selecta Math. {\bf 1} (1995) 347,
hep-th/9407025.

\bibitem{Lu:2003ze}
H.~Lu, C.~N.~Pope, K.~S.~Stelle and P.~K.~Townsend,
``Supersymmetric Deformations of $G_2$ Manifolds from Higher Order Corrections to String and M Theory,''
JHEP {\bf 0410}, 019 (2004), hep-th/0312002.

\bibitem{GHR}
S. J. Gates, Jr., C. Hull, and M. Rocek,  ``Twisted Multiplets And New Supersymmetric Nonlinear Sigma Models,''
Nucl. Phys. {\bf B248} (1984) 157.

\bibitem{NS}
D. Nemeschansky and A. Sen, ``Conformal Invariance Of Supersymmetric $\sigma$ Models On Calabi-Yau Manifolds,''
Phys. Lett. {\bf B178} (1986) 365.

\bibitem{EW}
E. Witten, ``New Issues In Manifolds Of $SU(3)$ Holonomy,'' Nucl. Phys. {\bf B268} (1986) 79.

\bibitem{DS}
M. Dine and N. Seiberg, ``Is The Superstring Weakly Coupled?'' Phys. Lett. {\bf B162} (1985) 299.

\bibitem{DStwo}
M. Dine and N. Seiberg, ``Nonrenormalization Theorems In Superstring Theory,'' Phys. Rev. Lett. {\bf 57} (1986)
2625-8.

\bibitem{DSW}
M. Dine, N. Seiberg, and E. Witten, ``Fayet-Iliopoulos Terms In String Theory,'' Nucl. Phys. {\bf B289} (1987) 589.

\bibitem{Kaste:2003zd} 
  P.~Kaste, R.~Minasian and A.~Tomasiello,
  ``Supersymmetric M theory compactifications with fluxes on seven-manifolds and G structures,''
  JHEP {\bf 0307}, 004 (2003),
  hep-th/0303127.

\bibitem{Bryant:2005mz}
R.~L.~Bryant,``Some Remarks on $G_2$-Structures'', arXiv:math/0305124.

\bibitem{kari}
S. Karigiannis, ``Flows of $G_2$-Structures, I'', arXiv:math/0702077.

\bibitem{Grigorian:2008tc}
S.~Grigorian and S.~-T.~Yau,
``Local Geometry of the $G_2$ Moduli Space'',
Commun.\ Math.\ Phys.\  {\bf 287}, 459 (2009), arXiv:0802.0723.

\bibitem{Grigorian:2009ge}
S.~Grigorian,
``Moduli Spaces of $G_2$ Manifolds'',
Rev.\ Math.\ Phys.\  {\bf 22}, 1061 (2010),  arXiv:0911.2185.


\bibitem{Lu:2004ng}
H.~Lu, C.~N.~Pope, K.~S.~Stelle and P.~K.~Townsend,
``String and M-Theory Deformations of Manifolds with Special Holonomy,''
JHEP {\bf 0507}, 075 (2005), hep-th/0410176.

\bibitem{joyce}
D.~Joyce, {\it Compact Manifolds with Special Holonomy}, Oxford Mathematical Monographs series by Oxford University Press.


\bibitem{kari1}
S. Karigiannis, ``Deformations of $G_2$ and $\Spin(7)$ Structures on Manifolds'',
Canadian Journal of Mathematics 57, 1012-1055 (2005), arXiv:math/0301218.

\bibitem{HM}
J. A. Harvey and G. Moore, ``Superpotentials
and Membrane Instantons,'' hep-th/9907026.

\bibitem{GP}
J. Gutowski and G. Papadopoulos,  ``Moduli Spaces And
Brane Solitons for M-theory Compactications on Holonomy $G_2$ Manifolds,''
Nucl. Phys. {\bf B615} (2001) 237-265, hep-th/0104105.

\bibitem{BW}
C. Beasley and E. Witten, ``A Note On Fluxes And Superpotentials In M-theory
Compactifications On Manifolds of $G_2$ Holonomy,'' hep-th/0203061.

\bibitem{Hitchin}
N. Hitchin, ``The Geometry Of Three-Forms In Six And Seven Dimensions,'' arXiv:math/0010054.

\bibitem{BeckerBecker}
K. Becker and M. Becker,  ``Supersymmetry Breaking, M-Theory, and Fluxes,''
JHEP {\bf 0107} (2001) 038, hep-th/0107044.

\bibitem{Becker:2000jc} 
  K.~Becker,
  ``A Note on compactifications on spin(7) - holonomy manifolds,''
  JHEP {\bf 0105}, 003 (2001),
  hep-th/0011114.

\end{thebibliography}
\end{document}